\documentclass[letter]{article}
\usepackage{jheppub}

\usepackage{etoolbox}
    \makeatletter 
    \patchcmd{\maketitle}{\@fpheader}{}{}{} 
\makeatother

\usepackage{array}
\usepackage{graphicx}
\usepackage{tabmac}
\usepackage{mathrsfs}
\usepackage{latexsym}
\usepackage{amsthm,amsopn,amsfonts,amssymb,epsfig,stmaryrd,color}
\usepackage[active]{srcltx}

\newlength{\eqboxstorage}

\def\fr{\frac}

\makeatletter
\renewcommand{\subsubsection}{\@startsection
{subsubsection}
{3}
{0mm}
{\baselineskip}
{-0.5\baselineskip}
{\normalfont\normalsize\bfseries}}
\makeatother

\theoremstyle{remark}

\def\la{{\lambda}}

\def\cal L{{\mathcal L}}

\def\Z{{\mathbb Z}}

\newcommand{\cercle}[1]{\ensuremath{\setlength{\unitlength}{1ex}\begin{picture}(2.8,2.8)\put(1.4,0.7){\circle{2.7}\makebox(-5.6,0){#1}}\end{picture}}}
\newcommand{\tcercle}[1]{\ensuremath{\setlength{\unitlength}{1ex}\begin{picture}(2.8,2.8)\put(1.4,1.4){\circle{2.7}\makebox(-5.6,0){#1}}\end{picture}}}

\let\d\partial
\let\n\noindent

\def\R{{\rangle}}
\def\L{{\rangle}}

\def\G{{\mathcal G}}

\def\B{{\mathcal B}}

\let\la\lambda
\let\La\Lambda
\let\Om\Omega

\let\ta\theta

\let\rw\rightarrow

\let\ti\tilde
\newcommand{\LL}{\ensuremath{\langle\!\langle}}
\newcommand{\RR}{\ensuremath{\rangle\!\rangle}}

\def\ic{{\mathrm {ic}}}
\def\rc{{\mathrm {rc}}}

\def\cd{{\circledast}}

\def\co#1{{\color{red}#1}}

\def\lrw{\leftrightarrow}

\let\l\left
\let\r\right
\let\e\epsilon

\let\a\alpha
\def\A{{\square}}
\def\L{{\mathcal L}}

\def\lev{{\rm level}}
\def\beq{\begin{equation}}
\def\eeq{\end{equation}}
\def\bea{\begin{align}}
\def\eea{\end{align}}

\begin{document}

\title{Superconformal field theory and Jack superpolynomials}

\author[a]{Patrick Desrosiers}

\affiliation[a]{Instituto de Matem\'atica y F\'{\i}sica, Universidad de
Talca, 2 norte 685, Talca, Chile.}
\emailAdd{desrosiers@inst-mat.utalca.cl}

\author[a]{Luc Lapointe}

\emailAdd{lapointe@inst-mat.utalca.cl}

\author[b]{Pierre Mathieu}
\affiliation[b]{D\'epartement de physique, de g\'enie physique et
d'optique, Universit\'e Laval,  Qu\'ebec, Canada,  G1V 0A6.}
\emailAdd{pmathieu@phy.ulaval.ca}

 \abstract{
We  uncover a deep connection between the $\mathcal{N}=1$ superconformal field theory in $2D$ and eigenfunctions of the supersymmetric Sutherland model
known as Jack superpolynomials (sJacks).  Specifically,
the singular vector at level $rs/2$ of the Kac module labeled by the two integers $r$ and $s$  { can be obtained explicitly} as a sum of
sJacks whose indexing diagrams are contained in a rectangle  with $r$ columns and $s$ rows. As a second compelling evidence for the distinguished status of the sJack-basis in SCFT,  we find
that the degenerate Whittaker vectors (Gaiotto states), in both the  Neveu-Schwarz  and Ramond sectors, can be expressed 
rather simply  in terms of sJacks.  As a consequence, we are able to reformulate the 
supersymmetric version of the (degenerate) AGT conjecture in terms of the combinatorics of sJacks. 
}




 \maketitle

\section{Introduction}
The Jack polynomials
provide representations of the states in Virasoro highest-weight modules.
This statement is to be understood in the context of the Feigin-Fuchs representation and the fact that free-bosonic modes can be represented by power-sum functions and their derivatives (see for instance  \cite{SSAFR}). 
In consequence,  
every state in the Fock space is associated to a symmetric polynomial and any basis of symmetric polynomials  such as the Jack polynomials 
can be used to describe a given state.

But the   Jack-polynomial basis displays remarkable features. 
It was indeed observed in \cite{MY} -- and later in \cite{AMOSa} -- 
 that the singular vector in a Kac module can be represented by a single Jack polynomial.  Explicitly, if the central charge is $c=13-6(t+t^{-1})$ then the  singular vector $|\chi_{r,s}\R^{{\rm Vir}}$ at level $rs$ in the Kac module with highest-weight state $|h_{r,s}(t)\R^{{\rm Vir}}$ (cf. eq. \eqref{paravir})   is associated to the Jack with parameter  $\alpha=t$ and indexed by an $s\times r$  rectangular diagram: $|\chi_{r,s}\R^{{\rm Vir}} \longleftrightarrow P_{(r^s)}^{(\alpha)}$.
This relation clearly singularizes the Jack basis. Recent works on the AGT conjecture \cite{AGT}
have also pinpointed the Jack polynomials basis   \cite{AY,MMS,Yan},
a point that is particularity clear in  theories whose underlying algebra is the Virasoro times a $u(1)$ algebra  \cite{Alba}.

The aim of this work is to show that the role played by the Jacks in CFT is performed by their $\mathcal{N}=1$ supersymmetric version in superconformal field theory (SCFT).

 The Jack polynomials  are eigenfunctions of the Sutherland Hamiltonian, 
  which is the trigonometric version of the Calogero-Moser-Sutherland Hamiltonian (up the contribution of the ground-state wavefunction). 
They depend upon a parameter $\a$ which is nothing but the 
inverse of the Sutherland coupling constant    $\beta$.
The Jack superpolynomials \cite{DLM1,DLM3} { (or sJacks for short --  also known as Jack polynomials in superspace}) 
are natural extensions of the usual Jack polynomials in that they are eigenfunctions of the supersymmetric generalization of the Sutherland Hamiltonian 
\cite{SS,BTW}.   As such, they depend upon both commuting and anticommuting variables.

We first  show, via a free-field representation of the super-Virasoro algebra, that any state in SCFT highest-weight modules can be represented by a linear combination of sJacks. 
{But at this level, the sJacks are not singularized yet: a state can be represented by a combination of any family of symmetric superpolynomials. The bottom line of this correspondence is that a state can be associated with a superpartition.}

 The  truly striking property that  distinguishes the sJack-basis is that the singular vectors in SCFT can be represented by simple combinations of sJacks. 
And this representation holds true for both the Neveu-Schwarz (NS) and the Ramond (R) sectors.
 To illustrate the drastic simplification that our  new singular-vector representation provides, let us note that the {sJack-representation of the } NS singular vector at level 33/2  contains  only 11 terms, while its expression in terms of the modes $L_n$ and $G_r$ is expected to contain as many terms as the number of independent vectors in the Verma module at level 33/2, which is 1687.

 The  particular combination of sJacks that appear in the expression of a singular vector is controlled by a  diagram whose shape is a rectangle. But in the supersymmetric context, the rectangle-rule is more intricate than in the usual case where  the rectangle selects a single Jack.
 To give an idea of the rule, we must recall that, in the same way as a Jack  
  is labeled by a partition,  an sJack 
  is labeled by a superpartition. A superpartition is characterized by two ordinary partitions.  When glued together in a certain way, the  diagrams associated to these two partitions need to fill the rectangle with $r$ columns and $s$ 
rows for the labels corresponding to the singular vector $|\chi_{r,s}\R^{{\rm}}$ at 
level $rs/2$.  We call such superpartitions self-complementary.

{The self-complementarity specifies} those terms that appear in the representation of a given singular vector. But what is rather astonishing -- and a priori totally unexpected -- is that we can also  obtain
 all the coefficients in the linear combination  of sJacks that represents the singular vector.  
In other words,  using the sJack-basis
we end up with a fully explicit expression  for all the singular vectors in SCFT.
The general formula is given in eq \eqref{generalformula}.

It is worth mentioning that there are two infinite sequences of singular vectors that do contain a single representative  sJack, namely those for which $r$ or $s$ is equal to 1 (see eqs \eqref{vs1s} and \eqref{vsr1}.
Interestingly, these are precisely the single instances for which an explicit mode-expression was known  (see \cite{BsA} and \cite{Watts} for the NS and R sectors, respectively).

 The recent work on the AGT conjecture has provided a novel CFT-intrinsic way of revealing the special status of the Jack-basis. Recall that in specializing the AGT conjecture to asymptotically free theories, Gaiotto \cite{Gai} was led to the consideration of Whittaker vectors, namely vectors not annihilated by $L_1$ (in the degenerate or matter-free case)  or not annihilated by $L_{1}$ nor $L_{2}$ (in the general case).
 In the degenerate case, an elegant expression for the Whittaker vector has been obtained in the Jack basis \cite{Yan} {(see also \cite{AY})}. This result is generalized here to the superconformal case. We thus end up with a beautiful closed-form polynomial representation for the SCFT  degenerate Whittaker vector in both the NS and R sectors.
{ We stress the remarkable simplicity of these expressions and at the same time their very non-trivial character in the sense that they are not immediate extensions of their Virasoro relatives. Again, we take this as a further confirmation of the  relevance  of the sJack basis in SCFT.}

 The article is concerned with the relationship between sJacks and the states in SCFT. But for those readers unfamiliar with the analogous connection -- in particular, in relation with singular vectors -- in the non-supersymmetric case, a brief summary is presented  in Appendix~A.
  The core of the article is organized as follows.
 In  Section 2, the notions of highest-weight representations of the superconformal algebra and Kac modules are briefly reviewed. Then, by means of the free-field representation, a correspondence is made between the states  in highest-weight modules and  symmetric superpolynomials. 
 The  SCFT-distinguished family of symmetric superpolynomials,
 the sJacks, are 
 introduced in Section 3. To complement the formal definition of these
 polynomials, some explicit expansion formulas are given in Appendix C.  
 In Section 4, the lowest-level singular vectors are re-expressed in the sJacks basis: these linear combinations of sJacks are seen to be both simple, in the sense that they involve very few terms, and tractable, in that their coefficients factorize nicely. By inspection, we  have obtained the full expression for 
 three infinite series, corresponding to the $r=1,2,3$ singular vectors. 
 Next, it is shown that the sJacks duality-property allows to relate the $(r,s)$  and $(s,r)$ singular vectors (a transformation rule that thereby produces three new infinite series: $s=1,2,3$).    These special  closed-form formulas are generalized in Section 5: there, the general expansion coefficient of a given sJack contributing to the polynomial representation of the $(r,s)$ singular vector is written in a fully explicit way. 
  The efficiency of the sJack basic in SCFT is further demonstrated in Section 6, where {are displayed} the explicit sJack-representations of the degenerate Whittaker vector {and its norm squared.   This leads, in the conclusion,  to a reformulation the supersymmetric analogue of the asymptotic AGT conjecture as a combinatorial conjecture involving scalar products of sJacks.}
 To  keep the presentation as simple as possible, only the NS sector is 
{ treated} in the main text. 
All the results related to the Ramond sector are relegated to Appendix B.

\section{Correspondence between SCFT and symmetric superpolynomials}

In this section, we present the connection between symmetric superpolynomials and states in highest-weight modules over the super-Virasoro algebra, which is given by the following commutations \cite{BKT,FQS}\begin{align}\label{svir}
&[L_n,L_m]=(n-m)L_{n+m}+\frac{c}{12}n(n^2-1)\delta_{n+m,0}\nonumber\\
&[L_n,G_k]=\left(\frac{n}{2}-k\right)G_{n+k}\nonumber\\
&\{G_k,G_l\}=2 L_{k+l}+\left(k^2-\frac{1}{4}\right)\frac{c}{3}\delta_{k+l,0},
\end{align}
where as before, $n,m\in \mathbb{Z}$ and $c$ denotes the central charge. In the NS sector,  
$k$ and $l$ belong to $\mathbb{Z}+1/2$, while in the R sector, $k,l \in\mathbb{Z}$.

 As in the Virasoro case (reviewed in Appendix \ref{JackVir}), this connection goes through the free-field representation of the conformal algebra. The relevant representation of the superconformal algebra is in terms of the modes of a free bosonic super-field, or equivalently, the modes of a free boson and a free fermion, a representation that is reviewed in Section \ref{ffrep} \cite{BKT}. The link with symmetric superpolynomials is presented in  Section \ref{Spower}, where the notion of superpartition is seen to  arise naturally. But before presenting this key relationship, we review the notion of Kac modules. 
 And, as mentioned in the introduction, in order to streamline the presentation, we confine ourself to the NS sector in the main text.

\subsection{Kac modules}

A highest-weight module over the NS sector is characterized by the central charge $c$ and a highest weight $h$,  and is spanned {over $\mathbb{C}$}
by all states of the form
\beq \label{descendant}G_{-k_1}\cdots G_{-k_p}L_{-n_1}\cdots L_{-n_q} |h\rangle^\mathrm{NS}\,\qquad k_i,n_i>0.\eeq 
{$|h\rangle^\mathrm{NS}$ is the highest-weight state, characterized by }
the following three conditions:
\begin{equation}\label{hwNS}  L_0|h\rangle^\mathrm{NS}= h |h\rangle^\mathrm{NS}, \qquad L_n|h\rangle^\mathrm{NS}=0 \quad\forall n>0\qquad\text{and}\qquad G_k|h\rangle^\mathrm{NS}=0\quad \forall r>0.\end{equation} 
Due to the third relation in \eqref{svir},  $L_{2k}$ and $L_{2k+1}$ can be respectively written as $G_kG_k$ and $\{G_k,G_{k+1}\}/2$, which implies that the first condition in \eqref{hwNS} is a consequence of the second one. Simple manipulations with \eqref{svir} show that requiring 
\beq \label{hwsg}L_0|h\rangle^\mathrm{NS}= h |h\rangle^\mathrm{NS}, \qquad  G_{\frac12}|h\R^\mathrm{NS}=0\qquad \text{and}\qquad G_{\frac32}|h\R^\mathrm{NS}=0 \eeq
suffices to establish the highest-weight nature of $|h\R^\mathrm{NS}$ in the NS sector.

Let us consider a state $|\chi\R^\mathrm{NS}$ at level $k$ in the highest-weight module just described, i.e.,  $|\chi\R^\mathrm{NS}$ is a linear combination of states of the form \eqref{descendant} such that $ L_0|\chi\R^\mathrm{NS}=(h+k)|\chi\R^\mathrm{NS} $.  We say that $|\chi\R$ is a singular vector (or null vector) if it behaves as a
 highest-weight state, meaning that it satisfies
\beq     G_{\frac12}|\chi\R^\mathrm{NS}=0\qquad \text{and}\qquad G_{\frac32}|\chi\R^\mathrm{NS}=0 . \eeq

As in the usual Virasoro case, there are special highest-weight states whose conformal dimension  $h=h_{r,s}$
is parametrized by two positive integers $(r,s)$ and is related to $c$ via the auxiliary parameter $t$:
\begin{align}\label{sch}
c=\frac{15}{2}-3\left(t+\frac{1}{t}\right)\qquad\text{and}\qquad h_{r,s}=\frac{t}{8}(r^2-1)+\frac{1}{8t}(s^2-1)-\frac{1}{4}(rs-1).
\end{align}
{This dimension formula pertains to the NS sector, in which $r+s$ is even. }  For such so-called Kac modules, there is a singular vector at level $rs/2$.\footnote{That  there is at most one singular vector at each level is proved in \cite{IK1}
 and the existence  of a singular vector at level $rs/2$ is demonstrated in \cite{KW}.}


 \subsection{Free-field representation}\label{ffrep}

  We introduce the  standard free bosonic and fermionic modes, denoted respectively  $a_n$ and   $b_r$, which satisfy
 \beq\label{cacao}
[a_n,a_m]=n\delta_{n+m,0},\qquad [a_0,\pi_0]=1,\qquad \{b_k,b_l\}=\delta_{k+l,0},
\eeq
{where 
 $\pi_0$ is the canonical conjugate of the zero-mode $a_0$. The indices $n,m$ are integer, while $k$ and $l$ are half-integer.
 We use the standard notation $:\, \, :$ for the normal ordering 
(the largest mode is placed at the right end, which introduces a sign if two fermionic modes are interchanged).}
It is a simple exercise to show that the following expressions provide  a representation of the NS-sector superconformal algebra  with central charge  $c=3/2-12\gamma^2$:
\begin{align}\label{ffLG}
L_n&=-\gamma (n+1)a_n+\frac{1}{2}\sum_{m\in\mathbb{Z}}:a_ma_{n-m}:+\frac{1}{4}\sum_{k\in\mathbb{Z}+\frac{1}{2}}(n-2k):b_kb_{n-k}: \nonumber
\\
G_k&=-2\gamma \left(k+\frac{1}{2}\right)b_k+\sum_{m\in\mathbb{Z}}a_mb_{k-m}.
\end{align}

{The Fock space $\mathscr{F}$ to be considered here is associated to  a free bosonic superfield. It is thus the tensor product of a free-boson Fock space 
 and that of a free fermion. 
For instance, the highest-weight states built on the vacuum state 
\beq
|0\R\equiv |0\R_{{\rm  bos}}\otimes|0\R_{{\rm fer}} ,\eeq are given by the one-parameter family 
 \beq|\eta\R\equiv |\eta\R_{{\rm  bos}}\otimes|0\R_{{\rm fer}},\qquad \text{with}\qquad |\eta\R_{{\rm  bos}}\equiv e^{\eta\pi_0} |0\R_{{\rm  bos}}\,,  \eeq
  Note that
 \beq
a_n|0\R=b_k|0\R=0, \; \quad \forall n,k>0.\eeq
  }Using the latter equation and  the commutation relation $[a_0,\pi_0]=1$, it follows that  \beq a_0|\eta\R=\eta|\eta\R.\eeq 
The full Fock space  $\mathscr{F}$   is the linear span over $\mathbb{C}$ of all monomials 
\beq b_{-k_1}\cdots b_{-k_p}a_{-n_1}\cdots a_{-n_q}|\eta\R ,\qquad k_i, l_i>0 \, . \eeq

 According to the representation \eqref{ffLG}, we have 
\beq L_0|\eta\rangle = (\tfrac12{\eta^2}-\gamma\eta) |\eta\rangle,\eeq
 which means that the conformal dimension of $|\eta\rangle$ is 
 \beq h=\tfrac12{\eta^2}-\gamma\eta.\eeq
 This last equation, together with $c=3/2-12\gamma^2$ and the parametrization \eqref{sch}, finally allows us to identify the  highest-weight state $|h_{r,s}\R^\mathrm{NS}$ of the Kac module with the Fock space state $|\eta_{r,s}\R$, where 
\begin{equation}
\label{seta}\eta_{r,s}=\frac{1}{2\sqrt{t}}\Big((1+r)t-(1+s)\Big). \end{equation}
Note that the correspondence also requires  
\beq \gamma=\eta_{0,0}=\frac{1}{2\sqrt{t}}(t-1).\eeq

\subsection{Correspondence between states and the power-sum superpolynomials}\label{Spower}

We introduce the $n$-th power-sum in infinitely many variables, denoted $p_n$ and its fermionic partner $\tilde p_n$ defined as \cite{DLM6}
\beq
  p_n=\sum_ix_i^n \quad(n>0)\qquad\text{and}\qquad\tilde p_n=\sum_i\ta_i x_i^{n},\quad (n\geq0)
\eeq
where $\ta_1,\theta_2,\ldots$ are anticommuting variables. Note that $\tilde p_0$ is non-trivial, being equal to $\sum_i\ta_i$.  Both $p_n$ and $\tilde p_n$ are instances of symmetric superpolynomials, that is, polynomials in the variables $x_1, x_2,\ldots$ and $\theta_1, \theta_2,\ldots$ that remain unchanged under any simultaneous permutation of the form $(x_i,\theta_i)\leftrightarrow (x_j,\theta_j)$.  An elementary result states that any element in  the vector space over $\mathbb{C}$ of all symmetric superpolynomials in infinitely many variables, denoted by  $\mathscr{R}$, can be uniquely written as a polynomial in $p_1,p_2,\ldots$ and $\tilde p_0, \tilde p_1,\ldots$ (see for instance  \cite{DLM6}).

We are now in position to formulate a correspondence between the free-field modes and the differential operators acting on $\mathscr{R}$:
\begin{align}\label{scor}
&a_{-n}\longleftrightarrow\frac{(-1)^{n-1}}{\sqrt{\alpha}}\, p_n\qquad 
&&a_n\longleftrightarrow  n(-1)^{n-1}\sqrt{\alpha} \,\frac{\partial}{\partial{p_n}}
\nonumber\\
&b_{-k}\longleftrightarrow \frac{(-1)^{k-1/2}}{\sqrt{\alpha}}\, \tilde p_{k-1/2}\qquad
&&b_k \longleftrightarrow (-1)^{k-1/2}\sqrt{\alpha}\, \frac{\partial}{\partial {\tilde p_{k-1/2}}} 
\end{align}
where $k,n>0$ and $\a$ is a non-zero free parameter.\footnote{Note that the first two relationships differ slightly from the corresponding  ones in the Virasoro case given in \eqref{precor}.}
The correspondence between $\mathscr{F}$  and $\mathscr{R}$ 
is given by
\begin{align} \label{cors}
|\eta\R&\longleftrightarrow 1\nonumber\\ 
a_0& \longleftrightarrow \eta\nonumber\\ 
b_{-k_1}\cdots b_{-k_m} a_{-n_1}\cdots a_{-n_p}\, |\eta\R& \longleftrightarrow\zeta\, \tilde p_{k_1-\frac12}\cdots \tilde p_{k_m-\frac12}\, p_{n_1}\cdots p_{n_p},
\end{align}
with
\beq\label{bounds}
 k_i> k_{i+i}\geq \tfrac12 \qquad\text{and}\qquad n_i\geq n_{i+1}\geq 1,\eeq
and $\zeta$ is a constant  described below.

The pattern of independent descendants states written on the lhs of \eqref{cors} is easily justified: we can obviously order all the fermionic modes at the left and in decreasing values since  they mutually anticommute and commute with the $a_n$'s. Relabeling the indices as
\begin{align} \label{indi}
 k_i-\tfrac12&= \La_i \quad  \quad\, \text{for}\quad 1\leq i\leq m,\nonumber\\
 n_i&=\La_{i+m}\quad \text{for}\quad  1\leq i\leq p=\ell-m,
 \end{align}
the inequalities \eqref{bounds} become
\begin{align}
&\La_{i}>\La_{i+1}\geq 0 \quad\text{for}\quad1\leq i\leq m-1,\nonumber\\
&\La_{i}\geq\La_{i+1}\geq 1 \quad \text{for}\quad m+1\leq i\leq \ell-1.
\end{align}

Thus, any state in $\mathscr{F}$ is in correspondence with a polynomial in $\mathscr{R}$  indexed by two partitions: $\La^a=(\La_1,\ldots,\La_m)$, whose elements are strictly decreasing   with $\La_m\geq 0$, and  $\La^s=(\La_{m+1},\ldots,\La_\ell)$, which is a standard partition with non-zero elements.  Together, the partitions $\La^a$ and $\La^s$ form the superpartition $\La=(\La_{1},\ldots,\La_m;\La_{m+1},\ldots,\La_\ell)$ \cite{DLM1}. The non-negative integer  $m$ is called the fermionic degree of the superpartition $\La$   while its bosonic degree is given by $|\La|=\sum_{i=1}^\ell\La_i$. If $|\La|=n$, we write $\deg(\La)=(n|m)$. Let us  also  mention that $\ell$ is called  the length of $\La$.

Superpartitions are fundamental objects in the theory of symmetric superpolynomials \cite{DLM6,DLM7,DLMeva}.  Any element in a basis of the space $\mathscr{R}$ is indexed by a superpartition.  For instance, if $f$ is a symmetric polynomial of degree $n$ in the variables $x_1,x_2,\ldots$ and degree $m$ in the variables $\theta_1,\theta_2,\ldots$, then we can always write
\beq f=\sum_\La c_\La p_\La \eeq
where the sum extends over all superpartitions $\La=(\La_{1},\ldots,\La_m;\La_{m+1},\ldots,\La_\ell)$ of degree $(n|m)$ and  where
$p_\La$ stands for to the following product of power-sums:
\beq 
\label{pdef}
p_\La=\tilde p_{\La_1}\cdots \tilde p_{\La_m}p_{\La_{m+1}}\cdots p_{\La_{\ell}} .\eeq

Having made the connection {\eqref{indi} between modes and} superpartitions, we easily see that the product of power-sums on the rhs of \eqref{cors} is nothing but the 
{polynomial $p_\La$ just defined in \eqref{pdef}.}  Moreover, we are now able to express the constant $\zeta\equiv \zeta_\La$ as
\beq
\zeta_\La=\frac{(-1)^{|\La|-\ell+m} }{\a^{\ell/2}}
\eeq
Let us rewrite the state on the lhs of the third relation in \eqref{scor} as
\beq \label{dab}
d_{-\La}=d_{-\La_1}\cdots d_{-\La_{\ell}}\qquad\text{with}\qquad
 d_{-\La_i}=\begin{cases}b_{-\La_i-\frac12}\quad \text{for}\quad \La_i\in\La^a\\
a_{-\La_i}\quad \quad\text{for}\quad \La_i\in\La^s.\end{cases}\eeq
In the Fock space, a generic term at level $k$ is a linear combination of states $d_{-\La}\, |\eta\R$ with $\lev(\La)=k$, the level -- the relative conformal dimension of the descendant state -- being read off the correspondence given in \eqref{dab}. 
Such states can thus be represented by linear combinations of the super-power-sum $p_\La$ of level $k$, i.e.,
\begin{align}\label{corlas}
\sum_{\lev(\Lambda)=k}c_\La \,d_{-\La}\, |\eta\R& \longleftrightarrow\sum_{\lev(\Lambda)=k}\tilde c_\La(\a)\, p_{\La} 
\end{align}
where $\tilde c_\La(\a)=\zeta_\La\, c_\La$.
The constraint on the sums ensures that all the terms have the same conformal dimension.  When expressed in terms of the superpartition data, it reads:
\begin{equation} \label{dlev}
\lev(\Lambda)=n+\frac{m}{2}\qquad\quad\text{if}\qquad\deg(\La)=(n|m)
\end{equation}
An important observation in that regard is that, at a given level $k$, the value of $m$ is not fixed: terms with different $m$ can appear. However, the above relation  forces the different values of $m$ to differ by even integers, {which guarantees the coherence of the equations with respect to their Bose-Fermi statistics}.

Note that the sum on the rhs of \eqref{corlas}
 can be rewritten in terms of any basis of $\mathscr{R}$.  
In particular, it can be rewritten in terms of the Jack superpolynomials that will be introduced in the next section.

\subsection{Differential representation of the super-Virasoro generators}\label{difrep}

The free field representation \eqref{ffLG} and the correspondence \eqref{scor} immediately imply that the generators $G_k$ and $L_n$  can be represented as differential operators acting on the space $\mathscr{R}$ of symmetric  superpolynomials. Let us denote these differential representations by $\mathcal{G}_r$ and $\mathcal{L}_n$ respectively. For instance, with the shorthand notation
\beq \label{shortnotation}\partial_n =\frac{\partial}{\partial p_n}, \qquad \tilde\partial_n =\frac{\partial}{\partial \tilde p_n}, \qquad \bar\eta =\sqrt{\alpha}\eta ,\qquad \bar\gamma =\sqrt{\alpha}\gamma \eeq
{the $osp(1,2)$-subalgebra generators are found to be represented by}
\begin{align}
&\mathcal{G}_{1/2}= (\bar\eta-2\bar\gamma) \ti\d_0 +\sum_{n>0}(n\,\ti p_{n-1}\, \d_{n} - p_n\, \ti \d_n)\label{g12} \\
&\mathcal{G}_{-1/2}=  \frac{\bar\eta}{\a} \tilde p_0 +{\sum_{n>0}(\,p_{n}\, \ti \d_{n-1} - n \ti p_n\, \d_n) }  \label{repG}
\end{align}  
and
\begin{align}
&\mathcal{L}_{1}= (\bar\eta-2\bar\gamma) \d_1 -{\sum_{n>0}\left((n+1)\, p_{n}\, \d_{n+1} +n\tilde p_{n-1}\, \ti \d_n\right) }  \nonumber \\
&\mathcal{L}_{0}=   \frac{1}{2\alpha}\bar\eta(\bar\eta-2\bar\gamma)  +\sum_{n>0}n\, p_{n}\, \d_{n} + \sum_{m\geq 0}\big(m+\frac{1}{2}\big)\, \ti p_{m}\, \ti \d_{m} \label{repL}  \\
&\mathcal{L}_{-1}=   \frac{\bar\eta}{\a}  p_1- \sum_{n>0}n\, p_{n+1}\, \d_{n} - \sum_{m\geq 0}\big(m+1\big)\, \ti p_{m+1}\, \ti \d_{m} \nonumber
\end{align} 
In the following, we will need mostly the expression of $\G_{\frac12}$ and that of $\G_{\frac32}$ given by
\begin{align} \label{g32}
\G_{\frac32}&=-(\bar\eta+2-2\a)\ti\d_1+\a\d_1\ti\d_0-\sum_{n\geq 2}n\,\ti p_{n-2}\,\d_n+\sum_{n>0}p_n\,\ti\d_{n+1} 
\end{align}

\section{Jack superpolynomials}

We have seen in the last section that any element of the superconformal algebra can be  represented explicitly as a differential operator in the power-sums $p_n$ and $\tilde p_n$.   Moreover, to each state in the Kac module over the superconformal algebra, we can associate an element of the space $\mathscr{R}$ of symmetric superpolynomials.  Given that the set of all possible products of power-sums,  $p_\La$, constitute a basis for $\mathscr{R}$,  any singular vector of the Kac module can be written as a linear combination of the form $\sum_\La c_\La p_\La$.  However, as will become clear in the next section, the power-sum basis is not convenient since it does not lead to  simple expressions for the singular  vectors.    In the next  subsections, we give a short introduction to the Jack superpolynomials (sJacks), which will prove to provide the best polynomial representation in SCFT.

\n {\bf Warning:} As we will discuss shortly, the Jack superpolynomials are orthogonal with respect to a natural scalar product on the space of symmetric superpolynomials.  However, it should be stressed that the superpolynomials constructed in \cite{DLM1} are not orthogonal. The correct orthogonal basis
is introduced in \cite{DLM3} and from then on the terminology ``Jack superpolynomials'' has been reserved to the latter.

\subsection{{Supersymmetric Sutherland model and Jack superpolynomials}}

The Jack superpolynomials $P^{(\a)}_\La$ \cite{DLM3} are natural extensions of the usual Jack polynomials in that they are eigenfunctions (with the contribution of the ground-state wavefunction  
factored out) of the Hamiltonian in the supersymmetric generalization of the  Sutherland model (also known as the trigonometric  Calogero-Moser-Sutherland model). 
The latter supersymmetric model \cite{SS,BTW,DLM1} describes the interaction on the unit circle of $N$ bosonic particles -- represented by $N$ commuting variables $x_1,\ldots,x_N$ -- together with their respective fermionic partners -- represented by $N$ anticommuting variables $\theta_1,\ldots,\theta_N$.
In \cite{SS}, a set of $N$ mutually commuting operators  containing the 
Hamiltonian was found.  
However,  in contrast with the non-supersymmetric case, these operators do not determine their common eigenfunctions uniquely, i.e., the eigenvalues are degenerate \cite{DLM1}. 
There is however a second set of bosonic conserved quantities, which disappear when all $\ta_i$ are set equal to 0.  By considering the second lowest-order representative of this  second set, we end up with a fully non-degenerate description \cite{DLM3}. 
{After some transformations, the  
 two defining eigen-operators can be taken to be}
\cite{DLMeva}:\begin{align} \label{eqD} 
 &D= \frac{1}{2}\sum_{i=1}^N  \alpha x_i^2\partial_{x_i}^2
+\sum_{1 \leq i\neq j \leq N}\frac{x_ix_j}{x_i-x_j}\left(\partial_{x_i}-\frac{\theta_i-\theta_j}{x_i-x_j}\partial_{\theta_i}\right) \\ \label{eqDelta}
& \Delta= \sum_{i=1}^N \alpha x_i\theta_i\partial_{x_i}\partial_{\theta_i}+
\sum_{1 \leq i\neq j \leq N}
\frac{x_i\theta_j+x_j\theta_i}{x_i-x_j}\partial_{\theta_i}.
\end{align}
Clearly, the {polynomial} eigenfunctions of these operators  depend upon $x_i$ and $\ta_i$: they are thus superpolynomials. In addition, they {can be chosen to be}  invariant under the interchange of $(x_i,\ta_i)\lrw (x_j,\ta_j)$. 
When $\ta_i=0$, $\Delta$ disappears and $D$ reduces to the operator whose eigenfunctions are the ordinary Jack polynomials. 

 In order to define  precisely the sJacks, two more objects are required.  We first need to introduce the symmetric supermonomials
\cite{DLM1}:
\beq m_\La={\sum_{\sigma\in S_N}}' \ta_{\sigma(1)}\cdots \ta_{\sigma(m)}\,x_{\sigma(1)}^{\La_1}\cdots x_{\sigma(N)}^{\La_N},
\eeq 
where the prime indicates a sum over distinct permutations of $S_N$.    
For instance, 
for $N=3$, 
\begin{align}
m_{(2;1,1)}&=\ta_1x_1^2 x_2 x_3+\ta_2x_2^2x_1 x_3+\ta_3x_3^3 x_1x_2 \, ,\\
m_{(2,1;1)}&=\ta_1\ta_2(x_{1}^2x_2-x_2^2x_1)x_3+\ta_1\ta_3(x_{1}^2x_3-x_3^2x_1)x_2+\ta_2\ta_3(x^2_{2}x_3-x_3^2x_2)x_1\, .
\end{align}
More examples are displayed in Tables 1 and 3 of \cite{DLM1}.

We also need a partial order that allows to compare superpartitions and that generalizes the usual dominance order $<$ on partitions.  For technical reasons, we postpone to Section~\ref{Sdia} the definition of the generalized dominance order, still denoted $<$.  What is important for the moment is the following: the supermonomial $m_\Om$ is said to be lower than $m_\La$ whenever $\Om$ and $\La$ have the same degree $(n|m)$  and $\Om<\La$.

We are now in position to give a unique characterization of the sJacks.  For each superpartition $\La$, the sJack $P_\La^{(\a)}$ is  the unique symmetric superpolynomial satisfying the unitriangularity condition
\beq \label{unitriang} P_\La^{(\a)} =m_{\Lambda} + \text{lower terms}\, , \eeq
and the double-eigenfunction condition
\beq \label{doubleeigen} D P_\La^{(\a)} = \varepsilon_\La(\a)P_\La^{(\a)} \quad\text{and}\quad  \Delta P_\La^{(\a)} = \epsilon_\La(\a)P_\La^{(\a)}\, ,\eeq
where  $\varepsilon_\La(\a)$ and $\varepsilon_\La(\a)$ denote the eigenvalues. As shown in \cite{DLM3}, the above definition together with the exact formula for the action of $D$ and $\Delta$ on the monomial basis directly lead to explicit bi-determinantal formulas for the expansion 
\beq \label{defcLaOm} P_\La^{(\a)} =\sum_\Om c_{\La\Om} m_{\Om} \, . \eeq 
For a given degree $(n|m)$ , the coefficient $c_{\La\Om}$ can be viewed as the entry labeled by $\La,\Om$ in the transition matrix from the sJack basis to the supermonomial basis.  
The simplest transitions matrices are given in Appendix \ref{TablesJacks}.

 A further remark  is in order.  Implicitly, the above definition for the sJacks depends on the number $N$ of variables $(x_i,\theta_i)$.  However,   it can be shown that the expansion \eqref{defcLaOm} in the monomial basis  is stable with respect $N$: suppose that $N$ is greater than the bosonic degree of the superpartition $\La$, then as $N$ increases, the coefficients $c_{\La\Om}$ in \eqref{defcLaOm} remain the same. In other words,  the sJacks remain well-defined  in the situation where the number of variables is infinite (in which case 
they are called functions rather than polynomials  but this distinction is irrelevant for our purposes).   
 But an infinite number of variables is precisely the situation at hand, the reason being essentially rooted in that there are infinitely different modes in the Fock spaces. 

\subsection{Relation with the power-sums}

  As we have seen earlier, the correspondence between states in the Fock modules and symmetric superpolynomials is more easily understood in terms of  power-sums   than in terms of monomials.  We thus need to rewrite the sJacks as polynomials in the power-sums.

We recall that each element in a basis of the space $\mathscr{R}$ of symmetric superpolynomials in infinitely many variables is indexed by a superpartition.  Both the sJacks { and the products of power-sums, defined in \eqref{pdef},  
provide bases}
 of  $\mathscr{R}$.   Consequently, we can write
 \beq \label{transmate}  P_\La^{(\a)}=\sum_\Om e_{\La\Om} \, p_\Om\quad\text{and}\qquad p_\La= \sum_\Om \ti e_{\La\Om} P^{(\a)}_\Om,
 \eeq
 for some expansion coefficients $e_{\La\Om}$ and $\ti e_{\La\Om}$. These coefficients can be explicitly calculated  using the known expansion \eqref{defcLaOm} and the simple transformation rules relating the $p_\Om$ and the $m_\La$'s.  Appendix C  contains all the coefficients $e_{\La\Om}$ for the superpartitions of bosonic degree less than 4.  

{From} the second relation in \eqref{transmate} and the $\mathscr{F}$--$\mathscr{R}$ correspondence given in \eqref{corlas}, we see that any state in a NS highest-weight module can be represented as a linear combination of sJacks:
 \begin{align}\label{corlasj}
\sum_{\lev(\Lambda)=k}c_\La \,d_{-\La}\, |\eta\R& \longleftrightarrow\sum_{\lev(\Lambda)=k} b_\La(\a)\, P^{(\a)}_{\La} ,
\end{align}
 for some constants $b_\La(\a)$. 

The power-sums-{basis provides an alternative definition of the sJacks 
that avoids reference} to the differential operators $D$ and $\Delta$, which depend upon the number $N$ of variables.
For this, we need the natural scalar product 
on $\mathscr{R}$,  which is defined in terms of the power-sums as
\begin{equation} \label{scap} \LL \, 
{p_\La} \, | \, {p_\Om }\, \RR_\alpha=(-1)^{\binom{m}2}\, \alpha^{{\ell}(\La)}\, z_{\La^s}
\delta_{\La,\Om}\,, \end{equation}
where $z_{\La^s} $ is given by
\begin{equation}  \label{zlam}
z_{\La^s}=\prod_{i \geq 1} i^{n_{\La^s}(i)} {n_{\La^s}(i)!}\, ,
\end{equation}
with $n_{\La^s}(i)$ being the number of parts in $\La^s$ equal to $i$.  It turns out that the sJacks are orthogonal with respect to the scalar product \eqref{scap}:
\begin{equation} \label{ortho} \LL \, 
P_\La^{(\a)} \, | \,P_\Om^{(\a)}\, \RR_\alpha=0 \quad\text{when}\quad \La\ne\Om
\end{equation}
Actually,  to define the sJacks, this orthogonality condition can be used  instead of the double-eigenfunction characterization presented above. To be clear, the sJacks can be defined uniquely from the two conditions \eqref{unitriang}  and \eqref{ortho}.   Note that the norm of $P^{(\a)}_{\Lambda}$ with respect to the above scalar product is computable and  will be given below.


\subsection{Diagrams for superpartitions}\label{Sdia}

A key step in our finding of the general formula for the sJack representation of the singular vectors 
was a precise diagrammatic characterization of the contributing  sJacks at a given level (this is the subject of Sect.~\ref{srect}). This relies on the diagram representation of the superpartitions, which we review here.

Recall that $\La=(\La^a;\La^s)$ and that parts of $\La^a$ are distinct. Let us rewrite $\La$ by reordering all its parts in non-increasing values and, in order to keep track of the origin of each part (either in $\La^a$ or $\La^s$), let us circle  those parts belonging to $\La^a$;  
if a part is repeated and one of them is circled, then
the circled copy comes first.
The various entries give the number of boxes in each row of the corresponding diagram and for those entries that are circled, we add a circle at the end of the row.
For instance, we have:
\beq \La
=(3,1,0;2,1)= (\cercle{3},2,\cercle{1},1,\cercle{0}) 
: \qquad {\tableau[scY]{&&&\bl\tcercle{}\\&\\&\bl\tcercle{}\\ \\
    \bl\tcercle{}}}.\eeq
    Now let $\La^\cd$ be the partition associated to the diagram of $\La$ but with circles replaced by boxes and 
    $\La^*$ be the partition obtained by discarding the circles. For the superpartition of the previous example, we have thus
\begin{equation} \label{exdia}
     \La^\cd:\quad{\tableau[scY]{&&&\\&\\&\\\\ \\ }} \qquad
         \La^*:\quad{\tableau[scY]{&&\\&\\ \\ \\ }} \;. \eeq
Clearly, the pair of partitions $(\La^\cd,\La^*)$ completely fixes the superpartition $\La$. Note  that with this notation in hand, the level given in \eqref{dlev} can now be expressed more elegantly: 
 \beq \label{deflev}
 \lev(\La)=\frac12(\,{|\Lambda^*|+|\La^\circledast|}\,).\eeq

It is in terms of this pair  $(\La^\cd,\La^*)$ that the dominance ordering for superpartitions (used implicitly in \eqref{unitriang}) is defined:
\begin{equation*} 
 \Omega\leq\Lambda \quad \text{iff}
 \quad \Omega^* \leq \Lambda^*\quad \text{and}\quad
\Omega^{\circledast} \leq  \Lambda^{\circledast} , 
\end{equation*}
where the order $\leq$ on partitions is the usual dominance ordering 
\cite{Mac}:{
\begin{equation}
\lambda \geq \mu \quad \iff \quad | \lambda|=|\mu|
\quad \text{and}\quad \lambda_1+ \cdots +\lambda_i \geq
\mu_1+ \cdots +\mu_i\text {~for all~} i.
\end{equation}
}

The two partitions $\La^*$ and $\La^\cd$ also enter naturally in the expression for the norm squared  
of $P^{(\a)}_\La$, denoted $j_\La(\a)$, and defined via
\begin{equation} \label{normsjack} \LL \, 
P_\La^{(\a)} \, | \,P_\La^{(\a)}\, \RR_\alpha=(-1)^{\binom{m}{2}}\, j_\La,
\end{equation}
for $\La$ of fermionic degree $m$.
Explicitly, $j_\La$ reads \cite{DLMeva,LLN}:
\beq  j_\La(\a)=\a^m\prod_{s\in\La}\frac{h^\uparrow_\La(s)}{h^\downarrow_\La(s)},\label{norm}
\eeq 
 where
 the upper and lower hooks $h^{\uparrow\downarrow}$ of a box $s$ are defined as  \cite{DLMeva}:
\begin{align}\label{defhook}
&h^\uparrow_\La(s)=l_{\La^\cd}(s)+\a(a_{\La^*}(s)+1)\nonumber\\
&h^\downarrow_\La(s)=l_{\La^*}(s)+1+\a\,a_{\La^\cd}(s).
\end{align}
In the above expressions, given the box $s=(i,j)$ 
($i$-th row and $j$-th column) of a partition $\lambda$, 
the quantities $a_\la(s)$ and $l_\la(s)$ are defined as follows \cite{Mac}:
\beq a_\la(s)=\la_i-j\qquad\text{and}\qquad l_\la(s)=\la_j'-i,\eeq
where $\la'$ stands for the conjugate of $\la$, obtained by interchanging rows and columns.

The expression for the norm will enter in the construction of the dual of a singular vector.  
Furthermore, a particular ratio of upper and lower hooks will be
a basic building block of the general formula for the singular vectors.

\section{Superconformal singular vectors as sJacks: simplest cases}

\subsection{Strategy }

Let us denote by
 $\G_k$, as we did in Section \ref{difrep},  
 the  representation of the operators $G_k$ 
 obtained by replacing in \eqref{ffLG} the  modes $a_n$ and $b_l$ by the corresponding power-sums or their derivatives via \eqref{scor}.  The operators  $\G_k$  depend upon the parameters $\alpha$, $\eta$, and $\gamma$.  We recall that $\gamma$ determines the central charge $c$ of the algebra via $c=3/2-12\gamma^2$.

Now, considering that we are interested in  singular vectors of Kac modules, whose central charge $c(t)$ and conformal dimension $h_{r,s}(t) $ are given by  \eqref{sch}, we must identify 
$\gamma$ with $(t-1)/{(2\sqrt{t})}$, $\eta$ with $\eta_{r,s}$ given in \eqref{seta}, and finally $\alpha$ with $t$.   From now on, we thus assume that $t=\a$ and consider that each Kac module is fully determined by the values of $r$, $s$, and $\alpha$.    

 
 Since the singular vector $|\chi_{r,s}\rangle ^\mathrm{NS}$ is a descendant that behaves as a highest-weight state, to identify the representative of $|\chi_{r,s}\rangle^\mathrm{NS}$ in the space $\mathscr{R}$ of symmetric superpolynomials, it suffices to verify the differential-operator version of the conditions \eqref{hwsg} on a non-trivial element $f\in \mathscr{R}$.  
We thus search linear combinations $\sum v_\La P^{(\a)}_\La$ of fixed level $rs/2=n+m/2$ 
{(recall that $n=|\La|$ and $m$ is the fermonic degree of $\La$)}
 satisfying
\beq \label{but}{ \G_{\frac12}\big(\sum v_\La P^{(\a)}_\La\big) =0\qquad\text{and}\qquad
\G_{\frac32}\big(\sum v_\La P^{(\a)}_\La\big)=0 }
\eeq
when 
\begin{equation}\label{setab}\eta_{r,s}=\frac{1}{2\sqrt{\a}}\Big((1+r)\a-(1+s)\Big),\qquad\text{and}\qquad  \gamma=\frac{(\a-1)}{2\sqrt{\a}}.\end{equation}
Note that since $r+s$ is even {in the NS sector}, $r$ and $s$ are either both even or both odd. In the former case, the level is integer, so that $m$ is necessarily even (the singular vector is bosonic), while in the latter case, $m$ is odd (and the singular vector is fermionic). We stress that the precise normalization of the solutions of the system \eqref{but} is irrelevant. 

The explicit expressions of $\G_{\frac12}$ and $\G_{\frac32}$ are given in \eqref{g12} and \eqref{g32} respectively.
To calculate the action of these operators on a given sJack $P^{(\a)}_\La$, one first needs to expand $P^{(\a)}_\La$ in the power-sum basis. For example, given that  $P^{(\a)}_{(0;)} $ is equal to  $p_{(0;)}=\ti p_0$, we get
\beq  \G_{\frac12} P^{(\a)}_{(0;)}= (\bar\eta+1-\a)\qquad \text{and}\qquad \G_{\frac32} P^{(\a)}_{(0;)}=0. 
\eeq
{From} this simple calculation, we identify a first singular-vector representation:  $P^{(\a)}_{(0;)}$ when $\bar\eta=\a-1$, that is, $\eta=\eta_{1,1}$. In other words, we have
\beq \label{vs11}|\chi_{1,1}\R^\mathrm{NS}=G_{-\frac12}|0\R^\mathrm{NS} \, \longleftrightarrow \, P^{(\a)}_{(0;)} .\eeq  
The next simplest example is 
\beq \label{Pex1}
P^{(\a)}_{(0;1)}=p_{(0;1)}-p_{(1;0)}=\ti p_0 p_1-\ti p_1,\eeq for which
\begin{align}
\G_{\frac12} (\ti p_0 p_1-\ti p_1)&
= (\bar\eta+2-\a)\, p_{(1)}
\\
\G_{\frac32} (\ti p_0 p_1-\ti p_1)&=
(\bar\eta+2-\a).
\end{align}
We see that the terms on the rhs vanish if  $\eta=\eta_{1,3}$.  We thus obtain a second singular vector representation: 
\beq \label{vs13}|\chi_{1,3}\R^\mathrm{NS} \, \longleftrightarrow \, P^{(\a)}_{(0;1)} .\eeq   
Proceeding similarly and using the fact that 
\beq \label{Pex2}
P^{(\a)}_{(1;)}=\frac1{(1+\a)}(\, p_{(0;1)}+\alpha \, p_{(1;)}\,),\eeq
 we find
\beq \label{vs31} |\chi_{3,1}\R^\mathrm{NS} \, \longleftrightarrow \, P^{(\a)}_{(1;)} .\eeq

\subsection{Sample NS singular vectors in the sJack basis}

The three examples just worked out are too simple to provide any clue concerning the general pattern of  the sJack-representation of the singular vectors at higher levels. At level 2 for example, there is only one possible singular vector, which can be found in the Kac module with $r=s=2$.  Simple manipulations  lead to 
\begin{equation}\label{vs22}
|\chi_{2,2}\rangle^\mathrm{NS} \longleftrightarrow P^{(\alpha)}_{(2)} +\frac2{(1+\alpha)}P^{(\alpha)}_{(1,1)}+ \frac4{(1+\alpha)}P^{(\alpha)}_{(1,0;)}.
\end{equation} 
At first sight, the last representation seems disappointing since it contains all possible superpartitions whose level $n+m/2$ is equal to 2.   However, as we will see shortly, level 2 is the only case where the sJack basis does not show any particular advantage.  Indeed, at level 5/2, we get two very simple representations:
\beq \label{vs15} |\chi_{1,5}\R^\mathrm{NS} \, \longleftrightarrow \, P^{(\a)}_{(0;1,1)}\qquad \text{and}\qquad |\chi_{5,1}\R^\mathrm{NS} \, \longleftrightarrow \, P^{(\a)}_{(2;)}. \eeq 
At level 3, there is no singular vector, but level 7/2 exhibits once again a remarkable simplicity: 
\beq \label{vs17} |\chi_{1,7}\R^\mathrm{NS} \, \longleftrightarrow \, P^{(\a)}_{(0;1,1,1)}\qquad \text{and}\qquad |\chi_{7,1}\R^\mathrm{NS} \, \longleftrightarrow \, P^{(\a)}_{(3;)}\eeq
 At level 4, there are two singular vectors:  
\begin{multline}  \label{vs24} 
|\chi_{2,4}\R^\mathrm{NS} \longleftrightarrow P^{(\a)}_{{(2,2)}
}+\frac2{(\a+1)}\,P^{(\a)}_{
{(2,1,1)}}+\frac{12}{(\alpha+3)(\alpha+2)}\,P^{(\a)}_{{(1,1,1,1)}} \\
+\frac4{(\a+1)} \,P^{(\a)}_{{(1,0;2)}}+\frac8{(\alpha+3)(\alpha+2)}\, P^{(\a)}_{{(1,0;1,1)}}
\end{multline}
\begin{multline}\label{vs42}
|\chi_{4,2}\R^\mathrm{NS} \longleftrightarrow \,P^{(\a)}_{(4)}+
\frac { 4}{ ( 3\alpha+1 ) } \,P^{(\a)}_{(3,1)} +
 \frac{2(\a+2)} { ( 2\alpha+1 )(\alpha+1 )  }P^{(\a)}_{(2,2)} \\
+\frac { 8}{(3\alpha+1)} \,P^{(\a)}_{(3,0;)}
+\frac { 16 }{ ( 2\alpha+1 )   }\,P^{(\a)}_{(2,1;)}.
\end{multline} 
These representations in the sJack basis contain 5 terms, but in the power-sum basis (or in any other ``classical'' basis), they would contain exactly 10 terms, which corresponds to the number of superpartitions at level 4 {(and the number of independent states in the NS Verma module)}. What we observed {by comparing the sJack-representations} at levels 2 and 4 is {an insight} of a more general phenomenon for $r,s>1$: 
{the greater is the level $rs/2$ of $|\chi_{r,s}\R^\mathrm{NS}$, the greater is the difference between the number of terms in the sJack-representation of the singular vector and  the total number of superpartitions at level $rs/2$. }

\subsection{NS singular vectors with $r=1$ or $s=1$}

The examples given in equations \eqref{vs11}, \eqref{vs13}, \eqref{vs31}, \eqref{vs15}, and \eqref{vs17}  suggest that there are two infinite sequences of singular vectors that can be represented by a single sJack.   Indeed, one can show that
\beq \G_{\frac12}P^{(\a)}_{(0;1^{k})}=(\bar\eta+k+1-\alpha)P^{(\a)}_{(1^{k})}\qquad\text{and}\qquad \G_{\frac32}P^{(\a)}_{(0;1^{k})}=(\bar\eta+k+1-\alpha)P^{(\a)}_{(1^{k-1})}\,,\eeq
where $1^k$ denotes the partition whose $k$ parts are equal to 1.  This implies that $P^{(\a)}_{(0;1^{k})}$ can be associated to a singular vector whenever $\bar\eta=\alpha-k-1$, which is equivalent to $\eta=\eta_{1,s}$ with $s=2k+1$.   For $r=1$ and $s$ odd, we thus have
\begin{equation}\label{vs1s}\boxed{ |\chi_{1,s}\rangle^\mathrm{NS} \longleftrightarrow P^{(\alpha)}_{\big(0;1^{\frac{s-1}{2}}\big)} }\eeq 
The dual version of the above singular vector is, with $r$ odd,
\begin{equation}\label{vsr1}
\boxed{ |\chi_{r,1}\rangle^\mathrm{NS}  \longleftrightarrow P^{(\alpha)}_{\big(\frac{r-1}{2};\,\big)}\quad }\eeq 
This identification follows from
\beq \G_{\frac12}P^{(\a)}_{(k;)}=\frac{(\bar\eta+1-(k+1)\alpha)}{ (1+k\a)}P^{(\a)}_{(k)}\qquad\text{and}\qquad \G_{\frac32}P^{(\a)}_{(k;)}=\frac{k\a(\bar\eta+1-(k+1)\alpha)}{(1+(k-1)\a)(1+k\a)}P^{(\a)}_{({k-1})}.\eeq
The singular vectors \eqref{vs1s} and \eqref{vsr1} are the only {NS} ones we found for which there is a representation in terms of a single $P^{(\a)}_\La$ for a generic value of $\a$. In both cases, $m=1$, so that the levels are respectively $s/2$ and $r/2$, as they should.

\subsection{NS singular vectors for $r=2$}


Equations \eqref{vs22} and \eqref{vs24} provide the simplest examples of singular vectors with $r=2$.  The next singular vector of this type appears at level 6:
\begin{multline}  \label{vs26} 
|\chi_{2,6}\R^\mathrm{NS} \longleftrightarrow P^{(\a)}_{{(2,2,2)}
}+\frac2{(\a+1)}\,P^{(\a)}_{
{(2,2,1,1)}}\\+\frac{12}{(\alpha+3)(\alpha+2)}\,P^{(\a)}_{{(2,1,1,1,1)}}
+\frac4{(\a+1)} \,P^{(\a)}_{{(1,0;2,2)}}+\frac8{(\alpha+3)(\alpha+2)}\, P^{(\a)}_{{(1,0;2,1,1)}}\\+\frac{120}{(\alpha+3)(\alpha+4)(\alpha+5)}\, P^{(\a)}_{{(1,1,1,1,1,1)}}+
\frac{48}{(\alpha+3)(\alpha+4)(\alpha+5)}\, P^{(\a)}_{{(1,0;1,1,1,1)}}
\end{multline}
Note that the coefficient $P^{(\a)}_{(2^{s/2})}$ is set to 1, a
a normalization compatible with that of \eqref{vs22} and \eqref{vs24}.
{By inspection, we obtained} the following  general expression: 
\beq \boxed{
|\chi_{2,s}\R^\mathrm{NS} \longleftrightarrow  \sum_{\ell=0}^{s/2} 
\frac{\prod_{i=1}^{\ell} (4i-2)}{\prod_{i=\ell}^{2\ell-1}(\a+i)} \, P^{(\a)}_{(2^{s/2-\ell}1^{2\ell})}+\sum_{\ell'=1}^{s/2} \frac{4\prod_{i=1}^{\ell'-1} (4i-2)}{\prod_{i=\ell'}^{2\ell'-1}(\a+i)}\,P^{(\a)}_{(1,0;2^{s/2-\ell'}1^{2\ell'-1})} }
\label{vs2s} \eeq 
with the understanding that if the upper bound is smaller than the lower one in the products, then the product is replaced by 1. This formula has been tested up to level 16, that is, for all $s$ even and {$\leq16.$ }

\subsection{NS singular vectors for $r=3$}

{Up to this point}, only one singular vector with $r=3$ has been given, namely 
$|\chi_{3,1}\R^\mathrm{NS}$  in \eqref{vs31}.   
Let us display a few more examples conveniently normalized:
{\small \begin{equation} \label{vs33}
|\chi_{3,3}\rangle^\mathrm{NS} \longleftrightarrow P^{(\alpha)}_{(1;3)}+{\frac { ( \alpha+2 )  ( 3\alpha+
1 ) }{ 2( \alpha+1 ) ^{3}}} P^{(\alpha)}_{(1;2,1)}-{\frac {
 ( 3\alpha+1 ) }{ ( \alpha+1
 ) ^{2}}} P^{(\alpha)}_{(2,1,0;0)}
\end{equation}
\begin{multline}
|\chi_{3,5}\rangle^\mathrm{NS} \longleftrightarrow
P^{(\alpha)}_{{(1;3,3)}}+{\frac { ( \alpha+2 )  ( 3
\alpha+1 )}{ 2( \alpha+1 ) ^{3}}}  P^{(\alpha)}_{{(1;3,2,1)}}
+{\frac { 3( \alpha+4 )  ( 3\alpha+1 ) }{ ( \alpha+2 ) ^{2} ( 2\alpha+3
 )  ( \alpha+1 ) }} P^{(\alpha)}_{{(1;2,2,1,1)}}\\-{\frac { ( 3\alpha+1
 ) }{ ( \alpha+1 ) ^{2}}} P^{(\alpha)}_{{(2,1,0;3)}}-
 {\frac { 3( 3\alpha+1 ) }{ ( 2\alpha+3
 )  ( \alpha+2 )  ( \alpha+1 ) }} P^{(\alpha)}_{{(2,1,0;2,1)}}
\end{multline}
\begin{multline}
|\chi_{3,7}\rangle^\mathrm{NS} \longleftrightarrow 
P^{(\a)}_{{(1;3,3,3)}}+{\frac { ( \alpha+2 )  ( 3
\alpha+1 ) }{ 2( \alpha+1 ) ^{3}}} P^{(\a)}_{{(1;3,3,2,1)}}+
{\frac { 3( \alpha+4 )  ( 3\alpha+1 )}{ ( \alpha+2 ) ^{2} ( 2\alpha+3
 )  ( \alpha+1 ) }}  P^{(\a)}_{{(1;3,2,2,1,1)}} \\ 
 +{\frac { 9( \alpha+6
 )  ( 3\alpha+1 )  ( \alpha+4 )  ( 
3\alpha+5 ) }{2 ( 3+\alpha ) 
^{3} ( 2\alpha+5 )  ( \alpha+1 )  ( 
\alpha+2 ) ^{2}}}P^{(\a)}_{{(1;2,2,2,1,1,1)}}
-{\frac { ( 3\alpha+1 ) }{ ( \alpha+1 ) ^{2}}} P^{(\a)}_{{(2,1,0;3,3)}}
\\ -{\frac {3 ( 3
\alpha+1 ) }{ ( 2\alpha+3 ) 
 ( \alpha+2 )  ( \alpha+1 ) }} P^{(\a)}_{{(2,1,0;3,2,1)}}
 -{\frac { 3( \alpha+4 )  ( 3\alpha+1 )  ( 3\alpha
+5 ) }{ ( 2\alpha+5 ) 
 ( \alpha+1 )  ( 3+\alpha ) ^{2} ( \alpha+2
 ) ^{2}}}P^{(\a)}_{{(2,1,0;2,2,1,1)}}
 \end{multline}
}
By inspection of these cases, we were led to the following formula.
\medskip\\
{\small
\fbox{\begin{minipage}{\linewidth}\begin{multline} \nonumber
|\chi_{3,s}\R^\mathrm{NS} \longleftrightarrow  \sum_{\ell=0}^{(s-1)/2} 
\prod_{i=1}^{\ell}\frac{i(3\a+2i-1)(\a+2i)}{(\a+i)^{2}(2\a+\ell+i)} \,P^{(\a)}_{(1;3^{(s-1)/2-\ell},2^\ell,1^\ell)}
\\ 
-\frac{(3\a+1)}{(\a+1)^{2}}\sum_{\ell'=0}^{(s-3)/2} \prod_{i=1}^{\ell'}\frac{i(3\a+2i+1)(\a+2i)}{(\a+i+1)^{2}(2\a+\ell'+i+1)} \, P^{(\a)}_{(2,1,0;3^{(s-3)/2-\ell'},2^{\ell'},1^{\ell'}) }
\end{multline} 
\end{minipage}}
}
where again, we set $\prod_{i=1}^{0}f(i)=1$.
 This explicit formula has been tested up to level 33/2, that is, for all $s$ odd {and $\leq 11$.}

\subsection{Relating the singular vector $|\chi_{r,s}\R$ to $|\chi_{s,r}\R$}

{In the previous two subsections, we have presented two explicit formulas for which $r\leq s$. We now show how to interchange the two labels, which would thereby amount to recover the cases $r\geq s$ from those with the inequality reversed.}

Interchanging the labels $r$ and $s$ in the expression for $h_{r,s}$ can be obtained by interchanging $t\rw 1/t$ -- see eq. \eqref{sch}. In the representation of the singular vectors, this appears to be slightly more complicated. Indeed, neither $|\chi_{2,2}\R^\mathrm{NS}$ in \eqref{vs22} nor $|\chi_{3,3}\R^\mathrm{NS}$ in \eqref{vs33} are invariant under $\a\lrw 1/\a$. Similarly, 
 $|\chi_{2,4}\R^\mathrm{NS}$ and  $|\chi_{4,2}\R^\mathrm{NS}$  are not related by such a simple transformation (cf. the expressions \eqref{vs24} and \eqref{vs42}).

The correct rule for transforming $|\chi_{r,s}\R$ to $|\chi_{s,r}\R$ involves the duality transformation of the $P^{(\a)}_\La$'s (see \cite{Mac,Stan} for the ordinary Jack polynomials  and \cite{DLM7} in the super-case). 
Up to an irrelevant multiplicative factor, the rule is {
\beq \label{rule}
\text{if}\quad |\chi_{r,s}\R^\mathrm{NS} \longleftrightarrow
\sum_{\La} v_\La(\a)\, P^{(\a)}_\La\quad\text{then}\quad 
|\chi_{s,r}\R^\mathrm{NS} \longleftrightarrow\l[\sum_{\La} v_\La(\a)\,\tilde\omega_\a\, P^{(\a)}_\La \r]_{\a\rw1/\a}
, \eeq }
where\footnote{The operator $\tilde\omega_\a$ is given by $\tilde\omega_\a=(-1)^{\binom{m}2}\hat\omega_\a$, where the later is defined in \cite{DLM7} -- cf. Sect. 6.1 there. The action of $\hat\omega_\a$ on $P_\La^{(\a)}$ is not the natural way of defining this duality operation. It is actually defined as a mild transformation on the free-field representatives, that is, on the power sums:
$$\hat\omega_\a(p_n)=(-1)^{n-1}\a\, p_n\qquad\text{and}\qquad \hat\omega_\a(\ti p_n)=(-1)^n\a\,\ti p_n.$$
It is somewhat amazing that, when lifted at the level of sJacks, it yields a transformation as simple as \eqref{duP}.}
\beq \label{duP}\tilde\omega_\a \,P^{(\a)}_\La =j_\La(\a) \,P^{(1/\a)}_{\La'}.\eeq 
{ In the last equation, $\La'$ stands for the conjugate of $\La$}: its diagram is obtained from that of $\La$ by interchanging rows and columns. Finally, $j_\La(\alpha)$ denotes the norm squared of $P_\La^{(\a)}$ defined in eqs \eqref{normsjack} and \eqref{norm}.

Let us consider as an example, the construction of $ |\chi_{4,2}\R^\mathrm{NS}$ out of $ |\chi_{2,4}\R^\mathrm{NS}$ whose expression is given in \eqref{vs24}. 
{The transformation of $ |\chi_{2,4}\R^\mathrm{NS}$ according to the rule \eqref{rule} yields:}
\begin{align}
&\bar  \jmath_{(2,2)}\,P^{(\a)}_{(2,2)'}+
\bar  \jmath_{(2,1,1)}\frac{2\a}{(\a+1)}\,P^{(\a)}_{ {(2,1,1)'}}+
\bar  \jmath_{(1,1,1,1)}\frac{12\a^2}{(3\alpha+1)(2\alpha+1)}\,P^{(\a)}_{{(1,1,1,1)'}}\nonumber
 \\ +\,&
\bar  \jmath_{(1,0;2)}\frac{4\a}{(\a+1)} \,P^{(\a)}_{{(1,0;2)'}}+
\bar   \jmath_{(1,0;1,1)}\frac{8\a^2}{(3\alpha+1)(2\alpha+1)}\, P^{(\a)}_{{(1,0;1,1)'}}\end{align}
where $\bar \jmath_\La$ stands for $j_\La(1/\a)$ and the different $j_\La(\a)$ are
\begin{align}
&j_{(2,2)}=\frac{(2\a+1)\a^2}{\a+2},\qquad 
\quad j_{(2,1,1)}=\frac{(\a+1)^2\a^2}{\a+3},\qquad 
j_{(1^4)}=\frac1{24}{(\a+3)(\a+2)(\a+1)\a}\nonumber\\ &
j_{(1,0;2)}=\frac{2(\a+1)^2\a^2}{\a+2},\qquad j_{(1,0;1,1)}=\frac12{(\a+1)(\a+2)\a^2}.
\end{align}
Substituting these expressions in the previous equation yields
{$[({\a+1})/{2 \a^2} ]\,\,|\chi_{4,2}\R, $ }
where the representation of $ |\chi_{4,2}\R^\mathrm{NS}$  is given in \eqref{vs42}.

\section{Superconformal singular vectors as sJacks: general formula}

In the previous section, we presented 
certain closed-form representations $\sum v_\La P^{(\a)}_\La$  of the NS singular vectors.   A first observation is that all the coefficients $v_\La$ factorize nicely into products and ratios of terms $(i\a+j)$ for nonnegative integers $i,j$. This suggests an underlying natural combinatorial description.  A second observation is that not all superpartitions of a given degree $rs/2$ do occur in the sum $\sum v_\La P^{(\a)}_\La$. For instance,  in the representation of $|\chi_{1,5}\R$, only $(0;1,1)$ contributes, so that the following superpartitions 
of level 5/2 are missing: $(2;0),\,(1;1),\,(0;2)$. 

  In the following, we give a precise characterization of 
the sJacks at level $rs/2$ that appear in the representation of the $(r,s)$ singular vector, and then present an explicit expression for the coefficients $v_\La$.

\subsection{Self-complementary superpartitions}\label{srect}

A severe restriction selecting the partitions labeling the polynomials representing the singular vectors was already observed in the non-supersymmetric case: 
 only the partitions of degree $rs$ whose diagram fills perfectly the rectangular shape   with $r$ columns and $s$ rows do contribute. Of course, this criterion singles out the partition $(r^s)$.

It turns out that a similar {but less restrictive} criterion selects the superpartitions that occur in the representation of the NS singular vector at level $rs/2$.  {Recall that a superpartition $\La$ is equivalent to a pair of 
partitions $\La^*$ and $\La^\cd$ (cf. Sect. \ref{Sdia}).} Now, the superpartitions appearing in the representation of $|\chi_{r,s}\R^\mathrm{NS}$ are level--$rs/2$ superpartitions for which 
 the complement of the diagram of $\La^\cd$ in the rectangle with $r$ columns and $s$ rows corresponds to the diagram of 
$\La^*$ rotated by $\pi$. 
 The set of such superpartitions 
will be denoted $\A_{r,s}$, and a superpartition belonging to $\A_{r,s}$
will be called $(r,s)$-self-complementary. 
Take for instance the level--15/2 superpartition $(2,1,0;2,1)$; 
the $(3,5)$-self-complementarity is illustrated in the following diagram,
where the diagram with thick frames correspond to that of $\La^*$
rotated by $\pi$ and where we have used the diagram of $\Lambda$ instead of that of $\Lambda^\cd$ (whose outer shapes are the same) to better illustrate the self-complementarity:
\beq
\mbox{\small
\tableau[scY]{&&\bl \tcercle{}\\&&\tf\\&\bl \tcercle{}&\tf\\&\tf&\tf
\\\bl \tcercle{}&\tf&\tf}\;.
}
\eeq
 $(2,1,0;2,1)$ is thus an element of $\A_{3,5}$.
The complete set of  $(3,5)$-self-complementary superpartitions is 
\beq
\A_{3,5}=\{(1;3,3),\,(1;3,2;1),\,(1;2^2,1^2),\,(2,1,0;3),\,(2,1,0;2,1)\}.
\eeq
 and their corresponding diagrams are 
\beq 
\mbox{\small
\tableau[scY]{&&\\&&\\&\bl \tcercle{} &\tf\\\tf&\tf&\tf\\\tf&\tf&\tf}\qquad
\tableau[scY]{&&\\&&\tf\\&\bl \tcercle{} &\tf\\&\tf&\tf\\\tf&\tf&\tf}\qquad
\tableau[scY]{&&\tf\\&&\tf\\&\bl \tcercle{} &\tf\\&\tf&\tf\\&\tf&\tf}\qquad
\tableau[scY]{&&\\&& \bl \tcercle{}\\&\bl \tcercle{}&\tf\\\bl \tcercle{} &\tf&\tf\\\tf&\tf&\tf}\qquad
\tableau[scY]{&&\bl \tcercle{}\\&&\tf\\&\bl \tcercle{} &\tf\\&\tf&\tf\\\bl \tcercle{} &\tf&\tf} \;.
}
 \eeq  
Similarly, the diagrams corresponding to the $(2,6)$-self-complementary superpartitions:
\beq \A_{2,6}=\{ (2^3),\,(2^2,1^2),\, (2,1^4),\,(1^6),\,(1,0;2^2),\,(1,0;2,1^2),\,(1,0;1^4)\},
\eeq  are:
\beq \label{a26}
\mbox{\small
\tableau[scY]{&\\&\\&\\\tf&\tf\\ \tf&\tf\\ \tf&\tf}\qquad
\tableau[scY]{&\\&\\&\tf\\&\tf\\\tf&\tf\\\tf&\tf}\qquad
\tableau[scY]{&\\&\tf\\&\tf\\&\tf\\&\tf\\\tf&\tf}\qquad
\tableau[scY]{&\tf\\&\tf\\&\tf\\&\tf\\&\tf\\&\tf}\qquad
\tableau[scY]{&\\&\\&\bl \tcercle{} \\\bl \tcercle{} &\tf\\ \tf&\tf\\ \tf&\tf}\qquad
\tableau[scY]{&\\&\bl \tcercle{} \\&\tf\\&\tf\\ \bl \tcercle{} &\tf\\ \tf&\tf}\qquad
\tableau[scY]{&\bl \tcercle{}\\&\tf\\&\tf\\&\tf\\ &\tf\\ \bl \tcercle{} &\tf} \;.}\eeq 
 
That only Jack superpolynomials whose superpartition belongs to $\A_{r,s}$ appear in the representation of $|\chi_{r,s}\R$ has been verified for all our examples (which reach level 25/2). This rule explains the non-occurrence of the
level--5/2 superpartitions $(2;0),\,(1;1),\,(0;2)$ in the representation of $|\chi_{1,5}\R$ since they are not  $(1,5)$-self-complementary. Actually, $\A_{1,5}$, and more generally $\A_{1,s}$, contains a single element and the same is true for $\A_{r,1}$.

{Finally, it should be clear that $\La\in \A_{r,s}$ implies that $\La'\in \A_{s,r}$, a property that was implicitly assumed in the duality transformation of the previous subsection.}

\subsection{General formula for the coefficients of the sJack-form of the NS singular vectors}
In this section, we present the explicit form of the coefficient $v_\La$ in the representation of
\beq \label{explform}
|\chi_{r,s}\R^\mathrm{NS}\longleftrightarrow\sum_{\La\in\A_{r,s}} v_\La P_\La^{(\a)}.
\eeq
We stress that $\La\in\A_{r,s}$ automatically implies that $\lev(\La)=rs/2$. 
Let us fix at once our choice of normalization. For $r$ and $s$ even, the coefficient of the sJack labeled by $r^{s/2}$ is set equal to 1. Similarly, for $r$ and $s$ odd, the coefficient of the term with
$\La=((r-1)/2 ; r^{(s-1)/2})$ is fixed to be  1. All our previous examples are compatible with this choice.

{Note that, although the normalization of a singular vector
 is arbitrary, it turns out that by choosing the normalization appropriately
 was a key step in our finding of the general expression of the relative coefficients, because this particular choice
  reveals a  clear recursive pattern. 
  This pattern is manifest in the formulas for $r=2$ and $r=3$: changing $s\rw s+2$ produces extra terms but the coefficients of the terms that were present (e.g., terms with the same value of $\ell$ or $\ell'$ in \eqref{vs2s})
   are not affected.}

Before displaying the general  form of $v_\La$, we first need to introduce some notation.


\subsubsection{Introducing some combinatorial data}

We first identify two subsets of corners in a diagram  {\cite{SSAFR}}.
The {\it removable corners} (resp. {\it addable corners}) of a diagram associated to a partition 
are the boxes
that can be removed (resp. added)
in such a way that the resulting diagram is still associated to a partition.
The {\it interior corners} of a partition $\lambda$ are the squares $(i,j)$
such that the cell $(i+1,j+1)$ is an addable corner of $\lambda$.  
 For instance, in $\lambda=(8,4,3,3,1,1)$, we have marked the
interior corners by an i and the removable corners   by an r:
\beq
\mbox{\small
{\tableau[scY]{& & & \rm i & & & & \rm r \\ &&\rm i &\rm r \\&&\\\rm i && \rm r\\ \\  \rm r}}
}
\eeq

Next, we introduce a  notion intrinsic to self-complementary superpartitions, namely that of {\it complementary circles}, together with some associated data.
Let $\Lambda \in \square_{r,s}$.
If $\Lambda$ has a circle in row $k \leq \lfloor s/2\rfloor$, then it also has one in row $s+1-k$
by complementarity.  
Indeed, the presence of a circle in row $k$ implies $\La^\cd_k-\La_k^*=1$, while 
the self-complementarity forces 
\beq \La^\cd_k+\La^*_{s+1-k}=r
\qquad\text{and}\qquad \La^\cd_{s+1-k}+\La^*_{k}=r.
\eeq
Hence  $\La^\cd_{s+1-k}-\La^*_{s+1-k}=\La^\cd_k-\La^*_k=1$, which
means that there is a circle in row  $s+1-k$.
In this case, we say that the circles in
rows $k$ and $s+1-k$ are {\it complementary}, and denote the pair as $(u,d)$,
where $u$ and $d$ are the coordinates of the cells corresponding to 
the circles in rows
$k$ and $s+1-k$ ($u$ and $d$ obviously stand  respectively
for up and down).  
When the number of circle is odd, there will remain an unpaired circle in 
row $(s+1)/2$ ($s$ is then necessarily odd) which will play no role in the description of the general formula.  Consider for example
$\Lambda=(4,2,0;4,1) \in \square_{5,5}$:  
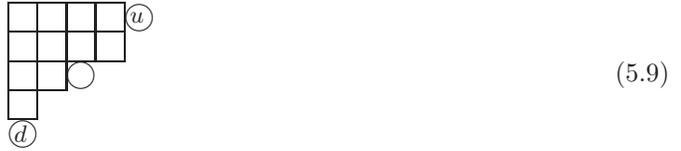
\beq\label{ex1}
\mbox{\small
{\tableau[scY]{ &  & & & \bl\tcercle{$u$}  \\ && &\\ & & \bl \tcercle{} \\  \\ \bl \tcercle{$d$} }}
}\eeq
We have only one pair $(u,d)$ of complementary circles given by $u=(1,5)$ and 
$d=(5,1)$. 
Let us denote by
 $\Lambda_{(u,d)}$  the superpartition
obtained by deleting the circle in position $u$ and changing the circle in position $d$ into a square. For the above example,  $\Lambda_{(u,d)}$ is the diagram
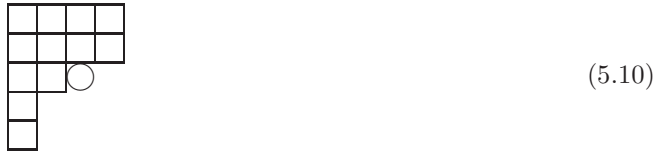
\beq\label{ex11}
\mbox{\small
{\tableau[scY]{& & &   \\ && &\\ & & \bl \tcercle{} \\  \\ \\}} }
\eeq
We also let $\Lambda^\circledast_{\rc(u,d)}$ 
(resp. $\Lambda^\circledast_{\ic(u,d)}$)
be the
set of removable (resp. inner) corners of $\Lambda^\circledast$ that lie 
 strictly between
cells $u$ and $d$.  
For the diagram \eqref{ex1} the interior (i)
and removable (r) corners of $\Lambda^{\circledast}$  strictly between $u$ and $d$
are
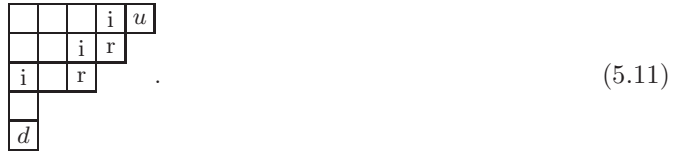
\beq\label{ex111}
\mbox{\small
{\tableau[scY]{& & & \rm i & u  \\ && \rm i& \rm r\\ \rm i & & \rm r\\  \\d }}.}
\eeq
Finally, given two cells $t_1=(i_1,j_1)$ and $t_2=(i_2,j_2)$
such that  $t_1$ is southwest of $t_2$,
we let 
\beq \|t_1,t_2\|= i_1-i_2 + \alpha(j_2-j_1)\eeq be the $\alpha$-distance between
$t_1$ and $t_2$.

\subsubsection{The explicit formula}
Let $(u_1,d_1),\dots,(u_{\ell},d_{\ell})$ be the pairs
of complementary circles in $\Lambda$ (which defines $\ell$), and for convenience let $\lambda=\Lambda^\circledast$.  Then, for $\La\in\A_{r,s}$, the coefficient of $P_\La^{(\a)}$ in the representation \eqref{explform} of 
$|\chi_{r,s}\R^\mathrm{NS},$ is
\beq \label{generalformula}
\boxed{v_{\Lambda} = {\small 2^{\ell} (-1)^{s\ell}  \ \prod_{j=1}^{\lfloor r/2 \rfloor} 
\left[ f(r+2-2j,2 \lambda_j' -s) 
\prod_{i=s+1-\lambda_j'}^{\lambda_j'} 
 \frac{{\bar h}_{\Lambda}^{\downarrow}(i,j)}{{\bar h}_{\Lambda}^{\uparrow}(i,j)} \right]
  \prod_{i=1}^\ell \left[ \frac{\prod_{t \in \Lambda^\circledast_{\rc(u_i,d_i)}} 
\|d_i,t \| }{\prod_{t \in \Lambda^\circledast_{\ic(u_i,d_i)}} \|d_i,t \| } \right] } }
\eeq
where  $f(p,q)$ stands for the product:
\beq
f(p,q) = 
\begin{cases}
\displaystyle{\prod_{j=1}^{q/2}\frac{p\a+2(j-1)}{(p-2)\a+2j}} & \text{if $p$ and $q$ are even,}\\
\displaystyle{\frac{(p-1)\alpha}{(p-1)\alpha+2} \prod_{j=1}^{(q-1)/2}\frac{p\a+2j-1}{(p-2)\a+2j+1}} & \text{if $p$ and $q$ are odd.}
\end{cases}
\eeq
with $f(p,0)=1$ and 
where
\beq
{{\bar h}^{\uparrow,\downarrow}_{\Lambda}(i,j)} = 
\begin{cases}
h^{\uparrow,\downarrow}_{\Lambda_{(u,d)}}(i,j) & \text {if a certain $d$ lies in column $j$}\\ 
h^{\uparrow,\downarrow}_{\Lambda}(i,j) & \text{otherwise,}
\end{cases}
\eeq
{where $ h^{\uparrow,\downarrow}_{\Lambda}$ are defined in \eqref{defhook}.}
Note that the product from
$s+1-\lambda_j'$ to $\lambda_j'$ is empty if $s+1-\lambda_j'> \lambda_j'$ (in which case it is replaced by 1).

We will illustrate the formula with the example
$\Lambda=(4,2,0;4,1) \in \square_{5,5}$, whose diagram is displayed in \eqref{ex1}.  
As already noticed, there is only one pair $(u,d)$ of complementary circles ($u=(1,5)$ and 
$d=(5,1)$), so that $\ell =1$; the first factor is thus $2 (-1)^s=-2$.  We have
$\lambda'=(\La^\cd)'=(5,3,3,2,1)$, and thus the $f$ factors are 
\beq f(5,5) = \frac{4\alpha}{(4\alpha+2)}  \frac{(5\alpha+1)}{(3\alpha+3)}  \frac{(5\alpha+3)}{(3\alpha+5)} \qquad {\rm and} \qquad f(3,1)=\frac{2\alpha}{(2\alpha+2)}\, .\eeq
 When $j=1$, since $\la_1'=5$ and $s+1-\la_1'=5+1-5=1$, the product of the ratio of hooks $h^\downarrow/h^\uparrow$ is over the boxes of the first column
in $\lambda_{u,d}$ displayed in \eqref{ex11} (we use the diagram of $\Lambda_{u,d}$ rather than that of $\La$  because these boxes belong to the
column of $d$).
This factor is 
\beq
\prod_{i=1}^{5} 
 \frac{{\bar h}^{\downarrow}(i,1)}{{\bar h}^{\uparrow}(i,1)} =
\frac{(3\alpha+5)}{(4\alpha+4)} \frac{(3\alpha+4)}{(4\alpha+3)} \frac{(2\alpha+3)}{(2\alpha+2)} \frac{2}{(\alpha+1)} \frac{1}{\alpha}.
\eeq
When $j=2$, since $\la_2'=3$ and $s+1-\la_2'=5+1-3=3$, the product is only on $i=3$ and thus the hooks are those of
cell $(3,2)$ in $\Lambda$ given by 
\beq  \frac{{\bar h}^{\downarrow}(3,2)}{{\bar h}^{\uparrow}(3,2)}= \frac{(\alpha+1)}{\alpha}.\eeq
Finally, the interior
and removable  corners of $\Lambda^{\circledast}$  between $u$ and $d$
are
indicated in \eqref{ex111}.
The product of their respective $\a$-distances to $d$ is
\beq
\frac{(2\alpha+2)(3 \alpha+3)}{2 (2\alpha+3)(3\alpha+4)}.
\eeq
If we make the product of all these terms, we obtain
\beq
v_{(4,2,0;4,1)} = -\frac{(5\alpha+1)(5\alpha+3)}
{(2\alpha+1)(\alpha+1)^2(4\alpha+3)} ,
\eeq
which coincides with our data.

One case that is not as trivial as it should be is the case of
$v_{(2;5,5)} \in \square_{5,5}$ which, by our choice of normalization,
is equal to 1.
In this case $\ell=0$ and so the only non-trivial part is in the first square
bracket.  The result is $f(5,1) f(3,1)$ times the hooks corresponding to
the boxes marked by  $\bullet$'s in the diagram
\beq
{\mbox{\small
\tableau[scY]{& & & &   \\&& & &\\ \bullet & \bullet& \bl \tcercle{} }}}
\eeq

But then one checks that the total product is equal to 1.
So in some sense the formula is not as optimal as it could be.

\subsubsection{Further illustration of the combinatorics underlying the general formula}

Here is an example based on a large diagram  that will
illustrating in more details 
the combinatorial structure underlying  the general formula.
Consider $\Lambda= (12,9,6,3,0;13,10,8,5,3) \in \square_{13,11}$.
The diagram of $\Lambda$, with its complementary pairs of circles
$(u_1,d_1)$ and $(u_2,d_2)$, is displayed on the left: 
\beq
\mbox{ \small
{\tableau[scY]{ & & & & & & & & & & &  &
\\ 
& & & & & & & & & & &  &\bl \tcercle{$u_1$} 
\\ 
& & & & & & & & & \\
& & & & & & & & & \bl \tcercle{$u_2$}\\
& & & & & & & \\
& & & & & &  \bl \tcercle{}\\
&& &&\\ & & &  \bl \tcercle{$d_2$} \\ && \\ \bl \tcercle{$d_1$} }}
\qquad\qquad{\tableau[scY]{ &  & & & & & & & & & &  &
\\ 
1 & & & & & & & & & & &  &\bl \tcercle{$u_1$} 
\\ 
1 & \bullet &  \bullet & & & & & & & \\
1 &  \bullet &  \bullet & 2 & & & & & & \bl \tcercle{$u_2$}\\
1 &  \bullet &  \bullet & 2 &  \bullet& & & \\
1 &  \bullet &  \bullet & 2 &  \bullet & \bullet &  \bl \tcercle{}\\
1 &  \bullet &  \bullet &2 &  \bullet \\ 
1 &  \bullet &  \bullet &  \bl \tcercle{$2$} \\
1 & \bullet &  \bullet \\
 \bl \tcercle{$1$} }}}\eeq
We have $\ell=2$ and thus the first factor is $2^2(-1)^{22}=4$.
 Since $\lambda'=(10,9,9,8,7,6,6,5,4,4,2,2,2)$, the product
of factors of type $f$ is $f(13,9) f(11,7) f(9,7) f(7,5) f(5,3) f(3,1)$.
The products of hooks will be those over the marked boxes in the above tableau at the right,
where the cells marked with a bullet are normal hooks, those marked
with a $1$ (including that in the circle) are hooks relative to
$\Lambda_{(u_1,d_1)}$,
while those marked
with a $2$ (including that in the circle) are hooks relative to
$\Lambda_{(u_2,d_2)}$,  where $\Lambda_{(u_1,d_1)}$ and $\Lambda_{(u_2,d_2)}$ are  given respectively by
\beq
\mbox{\small
\;{\tableau[scY]{ & & & & & & & & & & &  &
\\ 
& & & & & & & & & & &  
\\ 
& & & & & & & & & \\
& & & & & & & & & \bl \tcercle{}\\
& & & & & & & \\
& & & & & &  \bl \tcercle{}\\
&& &&\\ & & &  \bl \tcercle{} \\ && \\ & \bl }}
\quad {\rm and} \qquad
{\tableau[scY]{ & & & & & & & & & & &  &
\\ 
& & & & & & & & & & &  &\bl \tcercle{} 
\\ 
& & & & & & & & & \\
& & & & & & & & \\
& & & & & & & \\
& & & & & &  \bl \tcercle{}\\
&& &&\\ & & &   \\ && \\ \bl \tcercle{} }}
}\eeq
Finally, the interior and removable corners
of $\Lambda^\circledast$ between $u_1$ and $d_1$ and between $u_2$ and $d_2$ are
respectively
\beq
\mbox{\small
{\tableau[scY]{ & & & & & & & & & & &  &
\\ 
& & & & & & & & &  {\rm i}_7&&  &\bl \tcercle{$u_1$} 
\\ 
& & & & & & & & & \\
& & & & & & & {\rm i}_6& & {\rm r}_6 \\
& & & & & & {\rm i}_5&  {\rm r}_5\\
& & & & {\rm i}_4 & & {\rm r}_4\\
&& &{\rm i}_3&  {\rm r}_3\\ & & {\rm i}_2 &  {\rm r}_2  \\ {\rm i}_1 && {\rm r}_1 \\ \bl \tcercle{$d_1$} }}
\qquad\qquad{\tableau[scY]{ & & & & & & & & & & &  &
\\ 
& & & & & & & & &  &&  &\bl \tcercle{} 
\\ 
& & & & & & & & & \\
& & & & & & & {\rm i}_6& &\bl \tcercle{$u_2$} \\
& & & & & & {\rm i}_5&  {\rm r}_5\\
& & & & {\rm i}_4 & & {\rm r}_4\\
&& & {\rm i}_3&  {\rm r}_3\\ & &  & \bl \tcercle{$d_2$}   \\  &&  \\ \bl \tcercle{} }}
}\eeq
which correspond respectively  to the products 
\beq
\frac{\prod_{j=1}^6 \|d_1,{\rm r}_j \|}{\prod_{j=1}^7 \|d_1,{\rm i}_j \|}\qquad\text{and}\qquad
\frac{\prod_{j=3}^5 \|d_2,{\rm r}_j \|}{\prod_{j=3}^6 \|d_2,{\rm i}_j \|}.
\eeq
The final coefficient is the product of all these expressions.

\section{ sJack representation of the degenerate Whittaker vector}

As mentioned in the introduction, Gaiotto has showed that Whittaker vectors naturally appear in asymptotic limits of the AGT relations \cite{Gai}.  The simplest Whittaker vector is the degenerate one.  It is defined as the unique state $|h,g\R^\mathrm{Vir}$ in the Verma module over the Virasoro algebra, with highest weight state $|h\R^\mathrm{Vir}$, such that 
\beq \label{whittvir}  L_0  |h,g\R^\mathrm{Vir}= h  |h,g\R^\mathrm{Vir}, \qquad  L_1|h,g\R^\mathrm{Vir}=g \, |h,g\R^\mathrm{Vir},\qquad L_2  |h,g\R^\mathrm{Vir}=0\,,
\eeq
where $g$ is a formal parameter.  Note that according to the above definition, $ L_n  |h,g\R^\mathrm{Vir}=0$ for all $n\geq 2$.  An explicit representation of $|h,g\R^\mathrm{Vir}$ in terms of Jacks has been obtained in \cite{Yan} {(whose expression is a degenerate case of the one found in \cite{AY},  where the Whittaker vector of the deformed Virasoro algebra is related to the Macdonald polynomials)}.   

Here we find a closed-form  sJack-expansion for the supersymmetric degenerate Whittaker vector.  The latter have been recently studied in the context of the supersymmetric generalization  of the asymptotic AGT  relations in
 \cite{BF} for the NS sector.  

\subsection{Basic definitions}

Let us concentrate on the NS sector. We say that the bosonic state $|h,g\R^\mathrm{NS}$, in the Verma module with  highest weight state $|h\R^\mathrm{NS}$, is a degenerate Whittaker vector if it satisfies the following supersymmetric 
generalization of condition \eqref{whittvir}:
\beq \label{whittRS}  L_0  |h,g\R^\mathrm{NS}= h \, |h,g\R^\mathrm{NS}, \quad  G_{\frac{1}{2}}|h,g\R^\mathrm{NS}=\phi \sqrt{g} \,  |h,g\R^\mathrm{NS},\quad G_{\frac{3}{2}}  |h,g\R^\mathrm{NS}=0, \quad L_2  |h,g\R^\mathrm{NS}=0\,, 
\eeq
where $\phi$ is an auxiliary fermionic operator,  satisfying
$\phi^2=1$, introduced only for preserving the  coherence of the Bose-Fermi-statistics. Note that the above equations imply 
\beq   L_1  |h,g\R^\mathrm{NS}= g  |h,g\R^\mathrm{NS}, \qquad  G_{n+\frac{1}{2}}|h,g\R^\mathrm{NS}=0,\qquad L_n  |h,g\R^\mathrm{NS}=0\,, \qquad \forall \, n>1\,.
\eeq

For solving the system of equations \eqref{whittRS}, it is convenient to expand the Whittaker vector  level-by-level as
\beq
|h,g\R^\mathrm{NS} = \sum_{k\geq 0} g^{k} |h\R^\mathrm{NS}_{k}+ \phi \sum_{k\geq 0} g^{k+\frac{1}{2}} |h\R^\mathrm{NS}_{k+\frac{1}{2}}\,.
\eeq
where $|h\R^\mathrm{NS}_{k}$ denotes a descendant of {$|h\R^\mathrm{NS}\equiv |h\R_0^\mathrm{NS}$} at level $k$.  Then, the conditions \eqref{whittRS} can be translated into 
\beq \label{whittRSa}    G_{\frac{1}{2}}|h\R_k^\mathrm{NS}= |h\R_{k-\frac{1}{2}}^\mathrm{NS},\qquad G_{\frac{3}{2}}  |h\R_k^\mathrm{NS}=0, \qquad L_2  |h\R_k^\mathrm{NS}=0\,, \qquad \forall\, k\in \frac{\mathbb{N}}{2}.
\eeq

\subsection{Closed-form formula}
 
We now want to represent the  degenerate Whittaker vector at level $k$ as a symmetric superpolynomial:
\beq \label{eqwihittskacks}  \boxed{|h\R^\mathrm{NS}_{k} \longleftrightarrow W_k=\sum_{\lev ( \La)=k}w_\La P_\La }\eeq 
where we recall that $\lev ( \La)$ denotes the level of the superpartition $\La$ defined in \eqref{deflev}.   The coefficients are fixed by the following differential equations 
\beq  \mathcal{G}_{\frac{3}{2}}W_k=0 \qquad \mathcal{L}_2W_k=0\, ,  \qquad \forall \, k>0\, , \eeq 
  the recursion 
\beq \mathcal{G}_{\frac{1}{2}}W_k= W_{k-\frac{1}{2}},\qquad \forall \, k>0\, ,\eeq
and the initial condition $W_0=1$. 


We have found that the coefficients in \eqref{eqwihittskacks} 
can be written   in a rather simple  -- albeit nontrivial -- way as
\beq \label{coeffw}
 \boxed{ w_\La=\frac{(-1)^{\binom{m}{2}} \a^{|\La^*|}}{ \bar \eta}
 \left[\prod_{s\in \B\La}\frac1{h^\uparrow_\La(s)}\right]\,
\left[\prod_{ \substack{{(i,j)\in \La^\cd}\\ (i,j)\ne(1,1)}}\frac1{\bar\eta+i-\a j}\right]\, A(\La) }\eeq
using, as before, $\bar\eta=\sqrt{\a}\, \eta$, and where (using $\lambda=\Lambda^*$)
\beq\label{defA}
A(\La)= \prod_{\substack{(i,j)\in \la\\ (i,j) \text{is not a remo-}\\ \text{vable corner of } \la }}\frac{\bigl[2\bar\eta+1+i+\la_j'-\a(1+j+\la_i)\bigr]}{2\bar\eta+1+2i-\a(1+2j)}
\prod_{(i,j) \in \mathcal{F}\La} \frac{1}{2\bar\eta+1+i+\la_j'-\a(1+j+\la_i)}
 \eeq
In the last equation, 
the symbol $\mathcal{F}\La$ denotes the set of entries $s=(i,j)$ in the diagram of $\La$ that belong at the same time in a fermionic row and a fermionic column, where a row/column is said to be fermionic if it terminates with a circle.
In \eqref{coeffw}, $\mathcal{B}\La$ corresponds to the entries $s=(i,j)$ in the diagram of $\La$ that do not belong to $\mathcal{F}\La$ \cite{DLMeva}. 
 Note that when $m=0$, although the sJacks reduce to the ordinary Jacks, this is not so for the Whittaker vector (the defining conditions as well as the representation being different).

Let us illustrate the evaluation of $A(\La) $ by displaying some examples using the compact notation
$(2\bar\eta+a-\a b)\equiv(a,b)$. We  thus have
{\small 
\beq
{\tableau[scY]{&&\bl\tcercle{}\\ \bl\tcercle{}\\ }}
\;\rw\;A =\frac1{(3,3)};\qquad
{\tableau[scY]{& \\ \\ \bl\tcercle{}\\ }}
\;\rw\;A =\frac{(4,4)}{(3,3)};\qquad
{\tableau[scY]{&&\\ \bl\tcercle{}\\ }}\;\rw\;A =\frac{(3,6)}{(3,3)}
;\qquad{\tableau[scY]{&\bl\tcercle{}\\ \\ \\}}\;\rw\;A =\frac{(6,3)}{(3,3)}
\eeq }
We see that under conjugation, $(a,b)\rw(b,a)$. Here are few more examples:
{\small \beq
{\tableau[scY]{&& \bl\tcercle{}\\ & \bl\tcercle{}\\ }}\;\rw\;A =\frac{(4,4)}{(3,3)\,(3,5)};
\qquad {\tableau[scY]{&&\\ \\ }}\;\rw\;A =\frac{(3,6)\,(4,5)}{(3,3)\,(3,5)};
\qquad
{\tableau[scY]{&&\bl\tcercle{}\\& \\ }}
\;\rw\;A =\frac{(4,4)\,(4,5)\,(5,4)}{(3,3)\,(3,5)\,(5,3)}.
\eeq}

{We exemplify the general formula by displaying the explicit representation of the level 3 component of the degenerate Whittaker vector -- modifying slightly the above compact notation as
$(p\bar\eta+a-\a b)\equiv(a,b)_p$; it reads} {\small \begin{multline}
W_3=
{\frac { (3,6)_2   }{ 6(0,0)_1(1,2)_1(1,3)_1(3,3)_2
}} P^{(\a)}_{{(3)}}
+{\frac {
  \a (4,4)_2
}{ ( 2\a +1 ) (0,0)_1 ( 1,2 )_1 (2,1)_1
( 3,3)_2 }} P^{(\a)}_{{(2,1)}} \\ 
 +{\frac {   {\a }^{2} (6,3)_2 }{ ( \a +1 )  (\a +2 )(0,0)_1  (2,1)_1(3,1)_1  (3,3)_2}} P^{(\a)}_{{(1,1,1)}} 
  -{\frac {\a }{(0,0)_1(1,2)_1(1,3)_1 (2,1)_1(3,3)_2 }} P^{(\a)}_{{(2,0;)}}\\
 -{\frac { {
\a }^{2} }{ ( \a +1 ) (0,0)_1 ( 1,2)_1(2,1)_1 (3,1)_1(3,3)_2 }}P^{(\a)}_{{(1,0;1)}}
 \end{multline}
 }

\subsection{{Norm}}

The physical relevance of the degenerate Whittaker vector is that the square of its norm is equal to the degenerate limit of the four-point conformal block \cite{Gai,BF}  (see the conclusion for a further discussion of this point). It is thus of interest to compute the norm squared of this vector from the point of view of its polynomial representation. What we want { to} evaluate is
\beq\label{BFconj}  \big( \,|k \rangle \, , \, |k \rangle \, \big)_{c,h}\,, \eeq
where $|k\rangle\equiv|h \rangle_{k}^\mathrm{NS}$ denotes the level-$k$ degenerate Whittaker vector,  and where $( \,  \,,\, )_{c,h}$ stands for the usual Hermitian Shapovalov form   on the NS highest-weight module.  The latter form is non-zero and characterized by 
the following invariance property:
\beq  \label{invform} \big( \, G_r|\La \rangle \, , \, |{\Om} \rangle \, \big)_{c,h} = \big( \,|\La \rangle \, , \, G_{-r}|{\Om} \rangle \, \big)_{c,h}\eeq
for all half-integers $r$ and for all basic states $|\La \rangle,{|\Om \rangle}$ of the NS highest-weight module {($|\La \rangle$ is taken to be a shorthand for a state of the form \eqref{descendant} using the correspondence \eqref{indi})}.

We now want to rephrase the {norm squared \eqref{BFconj}} in the language of symmetric superpolynomials.  
Thanks to eqs \eqref{ffLG} and \eqref{scor}, we already know how to represent any element of the highest-weight module over the NS sector as an element of the space $\mathscr{R}$ of symmetric superpolynomials:
\beq G_{-r_1}\cdots G_{-r_p}L_{-n_1}\cdots L_{-n_q}|h\R^\mathrm{{NS}}\longleftrightarrow \mathcal{G}_{-r_1}\cdots \mathcal{G}_{-r_p}\mathcal{L}_{-n_1}\cdots \mathcal{L}_{-n_q}(1)=\sum_{\La} u_\La P_\La^{(\alpha)},\eeq
 where the sum runs over all  superpartitions whose  level is equal to $\sum_i r_i+\sum_j n_j$ and where the $u_\La$'s {denote} complex {coefficients} depending upon $\alpha$ and $\eta$.  We recall that these parameters are related to $c$ and $h$ of the module through
\beq c=\frac{3}{2}-12\gamma^2\, \qquad h=\frac{1}{2}\eta(\eta-2\gamma),\qquad \gamma=\frac{1}{2\sqrt{\alpha}}(\alpha-1)\,.\eeq

We also know that the space $\mathscr{R}$ is naturally equipped with the  scalar product $\LL \, \,| \,\, \RR_\alpha$ defined in \eqref{scap}.  {However,}  this scalar product is not  compatible with the invariance property \eqref{invform}.   Consider for instance the operators $\mathcal{L}_0$ and $\mathcal{G}_{\pm\frac{1}{2}}$ given in eqs \eqref{repG} and \eqref{repL}.  Then,  by letting $\beta$ be a function of $\alpha$ and by making use of (see \cite[eq (5.2)]{DLM7})
\beq  \LL \,p_n f \,| \, g\, \RR_\beta = \LL \,f \,| \, \beta n\partial_n g\, \RR_\beta \qquad \LL \,\tilde p_n f \,| \, g\, \RR_\beta =  \LL \,f \,| \, {\beta} \tilde \partial_n g\,  \RR_\beta\, , \eeq
one easily checks that  $\LL \, \mathcal{L}_{0}f\,| \, g\, \RR_\beta = \LL \, f\,| \, \mathcal{L}_{0}g\, \RR_\beta$. On the other hand, one also gets 
$  \LL \, \mathcal{G}_{\frac{1}{2}}f\,| \, g\, \RR_\beta \neq \LL \, f\,| \, \mathcal{G}_{-\frac{1}{2}}g\, \RR_\beta\, $  
whenever $\eta\neq \gamma$, which {prevents} the interpretation of the scalar product as an invariant form.

It is nevertheless possible to relate the scalar product $\LL \, \,| \,\, \RR_\alpha$ to the invariant bilinear form $( \,  \,,\, )_{c,h}$.  
{For this, we first make} a slight change of parametrization:
\beq \eta=\rho+\gamma \qquad \Longrightarrow \qquad h=\frac{1}{2}(\rho+\gamma)(\rho-\gamma).\eeq	
Let $\varphi$ denote the complex conjugation.  We set 
	\beq \varphi(\bar \rho)=-\bar \rho \qquad \text{and}\qquad \varphi(\bar \gamma)=\bar \gamma \, ,\eeq
where $\bar \rho=\sqrt{\alpha} \rho$ 
(recall that $\bar \gamma=\sqrt{\alpha} \gamma$).
We thus assume that $\bar \gamma$ (resp. $\bar \rho$) is real  (resp. {purely} imaginary), which implies that $\alpha$ is real.
Note that both $c$ and $h$, {being}  invariant under the action of $\varphi$,
must be real.  Then we define 
\beq  \LL \, f\,| \,g\, \RR^\varphi_\beta\equiv \LL \,\varphi(f) \,| \,g\, \RR_\beta, \eeq
which is a Hermitian form on $\mathscr{R}$.  Obviously,  $\LL \mathcal{L}_{0}f\,| \, g\, \RR^\varphi_\beta = \LL \, f\,| \, \mathcal{L}_{0}g\, \RR_\beta^\varphi$ for any $\beta$.  Moreover,
\beq \LL \, \mathcal{G}_{\frac{1}{2}}f\,\big| \, g\, \big\RR^\varphi_\beta - \LL \, f\,\big| \, \mathcal{G}_{-\frac{1}{2}}g\, \big\RR^\varphi_\beta=\left(1+\frac{\beta}{\alpha}\right)(\bar\rho-\bar \gamma) \LL \,\tilde \partial_0 f\,\big| \, g\, \RR^\varphi_\beta.
\eeq 
In other words, the invariance criterion requires  $\beta=-\alpha$.  Direct calculations then show that the latter condition is also sufficient, so that
\beq \label{invsesqui} \LL \, \mathcal{G}_{r}f\,\big| \, g\, \RR^\varphi_{-\alpha}= \LL \, f\,\big| \, \mathcal{G}_{-r} g\, \RR^\varphi_{-\alpha},\qquad \forall\, r\in\mathbb{Z}+\frac{1}{2},\qquad \forall\,f,g\in\mathscr{R}.
\eeq

Let us summarize what we have obtained so far.   We are given  two isomorphic vector spaces over $\mathbb{C}$:  1) the highest-weight module over the NS sector characterized by the central charge $c$ and  highest weight $h$ ; 2) the space $\mathscr{R}$ of symmetric superpolynomials whose coefficients are complex-valued rational functions of $\alpha$ and $\bar \rho$.  Both spaces are equipped with 
{a non-zero} invariant Hermitian form.  Moreover, $c$ and $h$ are real-valued functions of $\alpha$ and $\bar \rho$.  Now, as is well known (see for instance \cite[p.171]{Kirillov}), whenever $c$ and $h$ are real, the Hermitian Shapovalov form on the highest-weight module  is unique up to a multiplicative complex constant.  We can thus conclude 
that if $|\La\R \longleftrightarrow f $ and $|\Om\R \longleftrightarrow g$, then
\beq \label{equivforms}\big( \,|\La \rangle \, , \, |\Om \rangle \, \big)_{c,h} = z \,
\LL\, f \,| \, g \,\RR_{-\alpha}^{\varphi} \eeq
for some complex constant  $z$.

We finally return to the evaluation of \eqref{BFconj}. According to \eqref {eqwihittskacks}, the  degenerate Whittaker $|k\R$ is represented in $\mathscr{R}$ as
\beq W_k(\alpha,\bar \rho)=\sum_{\mathrm{level}(\La)=k} w_\La(\alpha,\bar \rho)P^{(\alpha)}_\La\,,\eeq
where $w_\La(\alpha,\bar \rho)$ is given by formula \eqref{coeffw}  with ${\bar \eta=\bar \rho+\bar \gamma}$.
Moreover, the equality \eqref{equivforms} allows us to write
\beq \big( \,|k \rangle \, , \, |k \rangle \, \big)_{c,h}= z \,{\LL} \, W_k(\alpha,\bar \rho)\,\big| \, W_k(\alpha,\bar \rho)\, {\RR}^\varphi_{-\alpha}=z \,{\LL} \, W_k(\alpha,-\bar \rho)\,\big| \, W_k(\alpha,\bar \rho)\, {\RR}_{-\alpha}\eeq
for some numeral constant $z$.    
Actually, in order to compare our results with those of \cite{BF}, it is more convenient to adjust the factor $z$ level by level:  we set
$z=4^{-\lfloor k\rfloor}$  
if the degenerate Whittaker vector is at level $k$.  Consequently,  
\beq\label{BFconj2}\boxed{ \big( \,|k \rangle \, , \, |k \rangle \, \big)_{c,h} 
=4^{-\lfloor k \rfloor }\,{ \sum_{\substack{\La,\Om\\\textrm{level}(\La)=\lev(\Om)=k}}  }
w_\La(\alpha,-\bar \rho)w_{\Om}(\alpha,\bar \rho)\,\LL\, P^{(\alpha)}_\La \,| \,P^{(\alpha)}_{\Om} \,\RR_{-\alpha}\, } .
\eeq
We have checked the concordance of the lhs of  \eqref{BFconj2} against all the  norms squared $( \,|k \rangle \, , \, |k \rangle \, )_{c,h} $  calculated  in \cite{BF} (i.e.,  up to level $k=5/2$).

Let us consider a nontrivial example: the 
evaluation of the norm squared for $k=3/2$.  By setting $\bar \eta=\bar \rho+\bar \gamma$ with $\bar \gamma= (\alpha-1)/2$ in eq \eqref{coeffw},  
we find 
\beq W_{\frac{3}{2}}(\alpha,\bar \rho)=\,{\frac {4}{ \left( 1-3\,\alpha+2\,\bar \rho \right)  \left(- 1
+\alpha+2\,\bar \rho \right) }} P^{(\alpha)}_{{(1;)}} +\,{\frac {4\alpha\,}{ \left( 
\alpha+1 \right)  \left( 2\,\bar \rho-\alpha+3 \right)  \left( -1+\alpha+2\,
\bar \rho \right) }} P^{(\alpha)}_{{(0;1)}}
\eeq 
{Using the decomposition of the above two sJacks in power sums, which are given in  eqs \eqref{Pex1} and \eqref{Pex2} (see also  Appendix C), we find}
\beq 
 \LL\, P^{(\alpha)}_{(0;1)} \,| \,P^{(\alpha)}_{(0;1)} \,\RR_{-\alpha} =\alpha(\alpha-1) \, , \quad 
{
\LL\, P^{(\alpha)}_{(0;1)} \,| \,P^{(\alpha)}_{(1;)} \,\RR_{-\alpha} =\fr{2\alpha^2}{(\alpha+1)} } \,,\quad
\LL\, P^{(\alpha)}_{(1;)} \,| \,P^{(\alpha)}_{(1;)} \,\RR_{-\alpha}=-\frac{\alpha^2(\alpha-1)}{(\alpha+1)^2}\,. \eeq
The rhs of \eqref{BFconj2} is thus equal to 
\beq {\frac { -4\left( -9{\alpha}^{2}+22\alpha-9+4{\bar \rho}^{2}
 \right) {\alpha}^{2}}{ 
 \left[ (1-\alpha)^2-4\bar \rho^2 \right]\,  
 \left[(3-\a)^2-4\bar \rho^2 \right]  
\,  \left[ (1-3\alpha)^2-4\bar \rho^2\right] }} = {\frac {{ c}+3{ h}}{8{ h}\, \left( { c}-9{ 
h}+2{ c}{ h}+6{{h}}^{2} \right) }}
\eeq
which in turn is exactly equal to the expression $\langle \frac32|\frac32\R$ in \cite{BF}.\footnote{Note that the notation used in \cite{BF} is slightly different from ours.  One goes from the former to the latter as follows: $\Delta\mapsto h$ and $c\mapsto 2c/3$.} 

\section{Conclusion}

Jack polynomials made an unexpected entry in CFT as a closed-form polynomial representation of singular vectors \cite{MY,AMOSa,SSAFR}. This is a fascinating, but still somewhat mysterious relationship.
One of our original motivations for studying the eigenfunctions of the supersymmetric Sutherland  model was precisely to generalize this connection at the SCFT level (cf. the concluding remarks in \cite{DLM1}.
This goal is achieved here. As plainly indicated in the main text, the supersymmetric case is more tricky since a linear combination of sJacks is required to describe a singular vector. Our most important technical result is the exact determination of the relative coefficients in this representation of a generic NS singular vector.

 It should be stressed that this general formula is provable; although its demonstration is a difficult problem, it is purely technical since we have all the ingredients at hand. Indeed, to demonstrate that a given combination of sJacks represents a singular vector, it suffices to show that it is annihilated by $\G_{\frac12}$ and $\G_{\frac32}$. For this, one needs the explicit action of these operators on the sJacks. Note that instead of $\G_{\frac32}$, one can use the expressions of $\G_{-\frac12}$ and  $\L_2$ given that their commutation generates the desired $\G_{\frac32}$.  Since we have derived in \cite{DLM0} the result of the action of  $\G_{\frac12}$ and $\G_{-\frac12}$ on $P_\La^{(\a)}$ via a set of Pieri-like rules, it remains to find the action of 
$\L_2$. {But $\L_2$ is essentially obtained from the commutation of $\partial_2$ and $2D+3\Delta$ (cf. \eqref{eqD} and \eqref{eqDelta}), where
the latter acts diagonally on the sJacks and the former can be obtained 
as (anti)commutators of operators whose action was 
given in  \cite{DLM0}.}

 Our second main result concerns the expression of the degenerate Whittaker vector, in both the NS and R sectors, in terms of a sum of sJacks with all the coefficients determined explicitly {(following the argument detailed
in the previous paragraph, we believe that this result is 
also provable with the tools we have)}.
Let us put this contribution in the context of the recent work revolving around the AGT conjecture.

The Jack-CFT correspondence 
 took a dramatic twist recently in works related to the AGT conjecture \cite{AGT}. Recall that this conjecture gives a relation between the conformal blocks in 2D CFT and 4D $\mathcal{N}=2$  supersymmetric gauge theories via the instanton part of the Nekrasov partition-function \cite{Nek}. 
{From} the CFT point of view, this provided a totally unexpected approach to the computation of conformal blocks that calls for an intrinsic CFT understanding. 
This issue was resolved in
\cite{Alba}, where by extending the Virasoro algebra by a $u(1)$ factor, a particular orthogonal basis of the extended algebra was obtained having the noteworthy property that the matrix elements 
of the primary fields in this basis
factorize in a very simple way -- precisely along the Nekrasov-partition-function pattern. Uncovering this basis essentially amounts to a proof of the $U(2)$-version of the AGT conjecture on the sphere. Interestingly,
this orthogonal basis diagonalizes a set of commuting integrals that turn out to be  the quantum Benjamin-Ono conserved quantities.

The basis vectors are specified by two diagrams.
 A remarkable fact is that whenever one diagram is void, the basis vector is the Jack labeled by the other diagram. 
 This is the starting point for a recursive procedure allowing the construction of the basis in the general case \cite{Alba}.

The supersymmetric generalization of the AGT conjecture was formulated in \cite{BF}: the supersymmetry on the CFT side corresponds to the replacement
of the $\mathbb{R}^4$ manifold of the gauge theory by the orbifold $\mathbb{R}^4/ \Z_2$. This proposal has been checked
 in the pure gauge case in \cite{BF}, and further confirmed in \cite{BBB}, when matter fields are present (for further discussion, see \cite{Ito,BMTa,coset}). 

In the pure-gauge case, the superconformal blocks are expressed in terms of the norm of the degenerate Whittaker vector. In \cite{BF}, this norm is calculated order-by-order, up to level 5/2. What we have obtained is a closed-form expression for this norm to all orders (cf. \eqref{BFconj2}).
We can thus  reformulate the correspondence proposed in \cite{BF} as a purely 
algebraic-combinatorial conjecture: 
\beq\label{BFconj3}\boxed{Z_{2k}\,\overset{}{=}        
4^{-\lfloor k \rfloor }\,{ \sum_{\substack{\La,\Om\\\textrm{level}(\La)=\lev(\Om)=k}}  }
w_\La(\alpha,\bar \rho)w_{ \Omega}(\alpha,-\bar \rho)\,\LL\, P^{(\alpha)}_\La \,| \,P^{(\alpha)}_{ \Omega} \,\RR_{-\alpha}\, } 
\eeq
where $Z_{2k}$ gives the $2k$ instanton-contribution to the partition function $Z^{-1}_{{\rm vec}}$.\footnote{The exact combinatorial formula for $Z_{2N}$ is given by the coefficient of $q^N$ on the LHS of eq (5.8) of \cite{BF} {with $Z_{{\rm vec}}$ given in  eq (5.6).}}  But the rhs of \eqref{BFconj3} is not tailor-made for a direct comparison with the Nekrasov partition function. Indeed, it is expressed in terms of the scalar product $\LL \, 
{P_\La^{(\a)}} \, | \, {P_\Om^{(\a)} }\, \RR_{-\a}$ (see \cite{AY,Yan} for analogous conclusions in other contexts).  Although the sJacks ${P_\La^{(\a)}} $ are orthogonal with respect to 
 $\LL \, |\, \RR_{\a}$, this is not so for  $\LL \, |\, \RR_{-\a}$. The expression of  $\LL \, 
{P_\La^{(\a)}} \, | \, {P_\Om^{(\a)} }\, \RR_{-\a}$ being unknown, this might not be a useful result at the end (it certainly raises the issue of trying to find convenient expansions for these quantities). But the potential usefulness  of the derived expression for the conformal block is beyond the scope of the present article, whose aim was to demonstrate the relevance of the sJack basis in SCFT. 
 
{We now expect that the role  played by the Jacks in the proof of the AGT conjecture will be performed by the sJacks in the supersymmetric version of the conjecture}.\footnote{This will be  most probably first verified in the $c=3/2$ (or $\a=1$) case, where the sJacks reduce to their Schur versions, generalizing the construction of \cite{BB} . }

\acknowledgments{This work was  supported by the Natural Sciences and Engineering Research Council of Canada; the
Fondo Nacional de Desarrollo Cient\'{\i}fico y
Tecnol\'ogico de Chile [\#1090034 to P.D., \#1090016 to L.L.]; and the Comisi\'on Nacional de Investigaci\'on Cient\'ifica y Tecnol\'ogica de Chile [Anillo de Investigaci\'on ACT56 Lattices and Symmetry].  P.D.\  wishes to thank Y.~Saint-Aubin for useful discussions on superconformal singular vectors and Shapovalov forms. }

\begin{appendix}

\section{Jack polynomials: basis and singular vectors in CFT}
\label{JackVir}

{The aim of this appendix is to review the relationship between Virasoro singular vectors and Jack polynomials.
As a preliminary step, we explain the statement made at the beginning of the introduction, where we  claim that
the Jacks
provide a representation of the  Virasoro states in highest-weight modules.
Such a representation is to be understood in the context of the free-field representation.}

The  free-field (or Feigin-Fuchs) representation of the Virasoro algebra  \cite{DF,CFT},
\beq [L_n,L_m]=(n-m)L_{n+m}+\frac{c}{12}n(n^2-1)\delta_{n+m,0},\eeq 
refers to the following expression of the Virasoro modes in terms of the free bosonic modes~$a_n$: 
\beq \label{Lvsa}
L_n =-\gamma (n+1)a_n+\frac{1}{2}\sum_{k\in\mathbb{Z}}:a_ka_{n-k}:\qquad\text{where}\quad [a_n,a_m]=n\delta_{n+m,0}.
\eeq 
In the last equation, the symbol $:\,:$ indicates the normal ordering  (the largest mode is placed at the right end) and $\gamma$ stands for a parameter related to the central charge via $c=1-12\gamma^2$.

The highest-weight states in the Fock space with vacuum $|0\rangle$ are given by  $e^{\eta\, \pi_0}|0\R\equiv|\eta\R$, 
where 
 $\pi_0$ is the zero-mode conjugated  to $a_0$
which means $[a_0,\pi_0]=1$, so that $a_0|\eta\R=\eta|\eta\R$.
The full module is obtained by acting in all possible ways with $a_n$, $n<0$.
Now, states in the Fock modules can be represented by symmetric functions. The basic relationships are (see e.g. \cite{AMOSa,SSAFR,CJ})
\begin{align} \label{cor}
|\eta\R&\longleftrightarrow 1\nonumber\\ 
a_0& \longleftrightarrow \eta\nonumber\\ 
a_{-\la_1}\cdots a_{-\la_\ell}\, |\eta\R& \longleftrightarrow \zeta'\, p_{\la_1}\cdots p_{\la_\ell} \qquad (\la_i\geq \la_{i+1}\geq 1)\end{align}
where $\zeta'$ is a constant and 
$p_n=\sum_{i\geq 1}x_i^n  $ is the standard $n$-th power-sum.
The precise correspondence between the modes $a_n$ and the polynomial $p_n$ will be taken to be
\begin{equation}\label{precor}a_{-n}\longleftrightarrow\frac{(-1)^{n-1}}{\sqrt{2\alpha} }\,p_n\qquad\text{and}\qquad
a_n\longleftrightarrow n (-1)^{n-1}\sqrt{2\alpha} \,\frac{\partial} {\partial{p_n}} \end{equation}
where $n>0$ and $\a$ stands for a free parameter. 

The sequence $(\la_1,\cdots,\la_\ell)$ in eq \eqref{cor} represents a partition  -- denoted $\la$ -- of length $\ell$. The  polynomial $p_{\la_1}\cdots p_{\la_\ell}$ in \eqref{cor} is the power-sum symmetric function indexed  by the partition $\la$ and written $p_\la$. 
The constant $\zeta'\equiv \zeta_\la$ is equal to $(-1)^{|\la|-\ell}(2\a)^{-\ell/2}$.
We could thus rewrite the state--polynomial correspondence \eqref{cor} as
$
a_{-\la} |\eta\R \longleftrightarrow \zeta_\la\,p_\la $
with $a_{-\la}=a_{-\la_1}\cdots a_{-\la_\ell}$. 
The sum $\sum_i \la_i=|\la|=n$ is called the degree of the partition. 


In the Fock space, a generic term at level $n$ is a linear combination of states $a_{-\la}\, |\eta\R$ with $|\la|=n$. Such states can thus be represented by linear combinations of the power-sum symmetric functions $p_\la$ of degree $n$, i.e.,
\begin{align}\label{corla}
\sum_{|\la|=n}c_\la \,a_{-\la}\, |\eta\R& \longleftrightarrow \sum_{|\la|=n} c_\la\,\zeta_\la \, p_{\la}
.\end{align}
The sum on the rhs \eqref{corla} can be rewritten in terms of any basis of symmetric functions, in particular, the Jack polynomials.

 Now, let us turn to the connection with singular vectors. A highest-weight module over the Virasoro algebra is characterized by the highest-weight state $|h\R^\mathrm{Vir}$ which satisfies the conditions $L_n| h\R^\mathrm{Vir}=0$ for $n>0$ and $L_0| h\R^\mathrm{Vir}=h|h\R^\mathrm{Vir}$. 
Kac modules are special highest-weight modules for which the conformal dimension 
$h\equiv h_{r,s}$ is related to the central charge $c$ via the following parametrization depending upon a {complex} number $t$:
\beq \label{paravir} c=13-6\left(t+\frac1{t} \right) \qquad \text{and}\qquad h_{r,s}=\frac14 ( {r^2}-1) {t} +\frac14( {s^2}-1)\frac1{ {t}} - \frac12( {rs}-1),\eeq 
where $ {r}$ and $ {s}$ are two positive integers. The remarkable feature of Kac modules is the occurrence of 
 a linear combination of the descendants (generated by the negative Virasoro modes) at level $n= {rs}$ that behaves like a highest-weight state: this is the  singular vector $|\chi_{r,s}\rangle^\mathrm{Vir}$. 
 The highest-weight state  $|h_{r,s}\R^\mathrm{Vir}$ characterizing the Kac module with $h_{r,s}$ and $c$ as in \eqref{paravir} can be  represented by the Fock state $|\eta_{r,s}\R$  if we choose 
\beq  h_{r,s}=\frac{\eta_{r,s}^2}{2}-\gamma \eta_{r,s},\qquad \eta_{r,s}=\frac{1}{{\sqrt{2t}}}\l((r+1)t-(s+1)\r)\qquad\text{and}\qquad \gamma=\eta_{0,0}.\eeq

The remarkable result of Mimachi and Yamada \cite{MY} (see also \cite{AMOSa, SSAFR,CJ}) is that the Jack polynomial indexed by the  rectangular partition $(r^s)$ (i.e., the partition with $s$ parts all equal to $r$) represents, up to a multiplicative constant,  the 
singular vector $|\chi_{r,s}\rangle^\mathrm{Vir}$:\footnote{The correspondence between the singular vectors for $c=1$ and the Schur polynomials is a much older result due to Goldstone (see page 322 in \cite{Segal}) and {generalized by} Wakimoto, Yamada \cite{WY}, and Wallach \cite{Wal}. }
\beq
|\chi_{r,s}\rangle^\mathrm{Vir}\longleftrightarrow P^{(\a)}_{(r^s)} \qquad(\text{when}\quad t=\a \quad\text{and}\quad \eta=\eta_{r,s})
.\eeq 

\section{The Ramond sector}\label{Rsector}

\subsection{Modules}
The Ramond (R) sector corresponds to the superconformal algebra \eqref{svir} in which the subscripts  $k$ and $l$ are integers.\footnote{For a thorough reference on the subject, see \cite{Dorr}. A more representation-theoretical treatment can be found in \cite{IK2}.}  Let $|h\R^\mathrm{R}$ denote the highest-weight vector of conformal dimension $h$, i.e., $L_{n}|h\R^\mathrm{R}=0=G_n|h\R^\mathrm{R}$ for all $n>0$ and $L_{0}|h\R^\mathrm{R}=h|h\R^\mathrm{R}$.  The highest-weight module {built from the state $|h\R^\mathrm{R}$ }
 can be defined as
\beq \mathscr{M}^\mathrm{R} = \mathrm{span}_\mathbb{C}\{\; G_{ -\La_1 }\cdots G_{ -\La_m }L_{-\La_{m+1}}\cdots L_{-\La_\ell} |h\R^\mathrm{R}\;\}
\eeq
where it is understood that 
\beq \label{eqspart} m\geq 0,\qquad \ell \geq 0,\qquad \La_1>\ldots> \La_m\geq 0,\qquad \La_{m+1}\geq \ldots \geq \La_{\ell}\geq 0\,.\eeq
 The last equation simply means that $\La=(\La_1,\ldots,\La_m;\La_{m+1},\ldots,\La_\ell)$ is a superpartition of fermionic degree $m$.     

The module  $ \mathscr{M}^\mathrm{R}$ is somewhat peculiar due to the relation 
\beq \label{eqG0} G_0^2=L_0-\frac{c}{24}\,.
\eeq
 The previous equation indeed implies that $|h\R^\mathrm{R}$ is supersymmetric only if $h=c/24$.  In what follows however, we restrict ourself to the more interesting case for which $h\neq c/24$.  Another consequence of \eqref{eqG0} is that $\mathscr{M}$ can be split into two submodules: 
\beq \mathscr{M}^{\mathrm{R},\pm} = \mathrm{span}_\mathbb{C}\{ G_{ -\La_1 }\cdots G_{ -\La_m }L_{-\La_{m+1}}\cdots L_{-\La_\ell} |h\R^\mathrm{R,\pm}\;\},
\eeq
where 
$ |h\R^\mathrm{R,+}=|h\R^\mathrm{R}$ and $|h\R^\mathrm{R,-}=G_0|h\R^\mathrm{R}$.  We say that the elements of $\mathscr{M}^{\mathrm{R},+}$ (resp.  $\mathscr{M}^{\mathrm{R},-}$) have a positive (resp. negative) chirality. 

An element $|\chi\R$ of  $\mathscr{M}$ is a singular vector if $G_n|\chi\R=0$ and $L_n|\chi\R=0$ for all $n>0$.    However, using the commutation relations among the $G_n$'s and $L_m$'s, {it is simple to check that the following two conditions are sufficient:} 
\beq  \label{condsingvecR} G_1|\chi\R=0\qquad\text{and}\qquad L_1|\chi\R=0.
\eeq
The algebra in the R sector  implies moreover that if $|\chi\R=\sum_\La c_\La G_{-\La^a}L_{-\La^s}|h\R$ is a singular vector in $\mathscr{M}^{\mathrm{R},+}$ then $|\chi^-\R=\sum_\La c_\La G_{-\La^a}L_{-\La^s}G_0|h\R$ is a singular vector in $\mathscr{M}^{\mathrm{R},-}$ , and conversely. In other words, the information about the singular vectors is exactly  the same for both submodules.  In order to avoid redundancies, from now on, we limit our study to the submodule with positive chirality.

We say that the highest-weight module $ \mathscr{M}^\mathrm{R,+}$  is a  Kac module over R (and of positive chirality) whenever the conformal dimension is related to the central charge  through the following parametrization:
\beq \label{hrsR} 
c=\frac{15}{2}-3\left(t+\frac{1}{t}\right)\qquad\text{and}\qquad h_{r,s}=\frac{t}{8}(r^2-1)+\frac{1}{8t}(s^2-1)-\frac{1}{4}(rs-1)+\frac{1}{16}.
\eeq
where $r$ and $s$ are positive integers such that $r-s$ is odd, while $t$ is a complex number. {Such a module is characterized by the presence of two singular vectors at level $rs/2$: a bosonic one, denoted $|\chi_{r,s}\R^{\mathrm{R}}$  and its  fermionic partner $G_0|\chi_{r,s}\R^{\mathrm{R}}$ (see for instance \cite[p. 616]{Dorr}).  Except for the {cases $r=1$ and $s=1$ \cite{Watts}, there are} no known  closed-form expressions for the singular vector $|\chi_{r,s}\R^{\mathrm{R}}$.  Below, we give new explicit formulas in the sJack basis.  

\subsection{{Free-field representation}}

We now turn our attention to the Fock-space representation of the R sector.  This requires the introduction of the following superalgebra involving the free-field modes $a_n$ and $b_n$, with $n\in\mathbb{Z}$, together with the vacuum charge operator $\pi_0$:
\beq \label{Ralgfock}
[a_n,a_m]=n\delta_{n+m,0} \, ,\qquad [a_0,\pi_0]=1\, ,\qquad \{b_n,b_m\}= \delta_{n+m,0}\,,
\eeq
all other commutations being zero.  Note in particular that 
$ b_0^2=1/2$.  
In other words, $\sqrt{2}b_0 $ is an involution in the superalgebra \eqref{Ralgfock}.  
Let $|0\R$ be the vacuum state, i.e.,  $a_k|0\R=0=b_k|0\R$ for all $k>0$.   We define a one-parameter family of highest-weight states as  $|\eta\R\equiv e^{\eta \pi_0}|0\R$. 
\beq a_0|\eta\R=\eta |\eta \R,\qquad a_n|\eta\R=0 \quad\text{and}\quad b_n|\eta\R=0,\qquad\forall\;n>0. 
\eeq
This allows us to introduce  the Fock space with highest weight $|\eta\R$ over the superalgebra \eqref{Ralgfock}:
\beq \mathscr{F} = \mathrm{span}_\mathbb{C}\{\; b_{ -\La_1 }\cdots b_{ -\La_m }a_{-\La_{m+1}}\cdots a_{-\La_\ell} |{\eta}\R\;\}
\eeq
where the labeling of the states  satisfies \eqref{eqspart}.  As one can check, the following equations yield a representation of the R sector on  $\mathscr{F}$:  
\begin{align} \label{ffrR} L_n&=-\gamma(n+1)a_n+\frac{1}{2}\sum_{k\in\mathbb{Z}}:a_ka_{n-k}:+ \frac{1}{4}\sum_{k\in\mathbb{Z}}\big(n-2k+\frac{1}{2}\big):b_kb_{n-k}:\\
 G_n&=-2\gamma\big(n+\frac{1}{2}\big)b_n+\sum_{k\in\mathbb{Z}}a_kb_{n-k} ,
\end{align}where the complex parameter $\gamma$ {(the background charge)}  is related to the central charge via $c=\tfrac{3}{2}-12 \gamma^2$.

\subsection{Correspondence with symmetric superpolynomials}

A correspondence between the  free modes just described and the differential operators acting on the space $\mathscr{R}$ of symmetric polynomials can be easily established.  A convenient choice is
\beq \label{corR}    a_n\longleftrightarrow \begin{cases} \frac{(-1)^{n-1}}{\sqrt{\alpha}}p_{-n} & n<0 \\ \eta & n=0\\ 
(-1)^{n-1} n\sqrt{\alpha}\partial_n &n>0\end{cases}\qquad  b_n\longleftrightarrow \begin{cases} \frac{(-1)^n}{\sqrt{2}}\tilde p_{-n} & n<0 \\ \frac{1}{\sqrt{2}}(\tilde p_0 +\tilde \partial_0) & n=0\\  (-1)^{n}\sqrt{2}\tilde \partial_n &n>0\end{cases}\, .
\eeq
The main differences between the above correspondence and that  of the NS sector \eqref{scor} 
{are the absence of $\a$ factors in the representation of the $b$ modes and}  the presence of the zero mode $b_0$ which is represented by a combination of {the fermionic polynomial $\ti p_0$} and its derivative.  Eq \eqref{corR} together with the identification  $|\eta\R\leftrightarrow 1$ induce  the following correspondence between $\mathscr{F}$ and $\mathscr{R}$:  
\beq \label{corR2} b_{-\La_1}\cdots b_{-\La_m}a_{-\La_{1+1}}\cdots a_{\La_\ell}|\eta\R\longleftrightarrow \zeta_\La p_\La\,, 
\eeq
where  
\beq \zeta_\La=\frac{(-1)^{|\La|-(\ell-m)}}{2^{m/2}\alpha^{(\ell-m)/2}} \, .\eeq
We stress that, in contradistinction with the NS sector, the level in the Fock space is now exactly equal to the bosonic degree of the superpolynomials. 

The free-field representation \eqref{ffrR}  and the correspondence \eqref{corR} yield a representation  of the super-Virasoro generators in  the R sector as differential operators acting on the space $\mathscr{R}$ of symmetric superpolynomials, for which sample expressions are
\begin{align}  \mathcal{L}_0&= \frac{1}{2}{\eta}(\eta-2\gamma)+\frac{1}{16}+\sum_{m\geq 1} m (p_m\partial_m+ \tilde p_m\tilde \partial_m) \nonumber\\
  \mathcal{G}_0&= \frac{1}{\sqrt{2}}(\eta-\gamma)(\tilde p_0+\tilde \partial_0) -\sqrt{\frac{2}{\alpha}}\sum_{m\geq 1} p_m\tilde\partial_m-\sqrt{\frac{\alpha}{2}}\sum_{m\geq 1}  m \tilde p_m \partial_m\nonumber\\
	 \mathcal{G}_1&=  {\sqrt{2}}(3\gamma-\eta) \tilde \partial_1+\sqrt{\frac{\alpha}{2}}(\tilde p_0+\tilde \d_0)\d_1+\sqrt{\frac{\alpha}{2}} \sum_{n\geq 1} n \tilde p_n\partial_{n+1}\sqrt{\frac{2}{\alpha}}\sum_{n\geq 1}   p_n \tilde \partial_{n+1}\nonumber\\
	 \mathcal{L}_1&=  {\sqrt{\alpha}}(\eta-2\gamma)   \partial_1- {\frac{1}{2}}(\tilde p_0+\tilde \d_0)\tilde \d_1-\sum_{n\geq 1} n  p_n\partial_{n+1}-{\frac{1}{2}}\sum_{n\geq 1}  (2n+1) \tilde p_n  \tilde \partial_{n+1}
	\end{align}
The complete differential representation induces the following correspondence between states of the module $ \mathscr{M}^\mathrm{R,+}$ with highest-weight  state  $|h\R^\mathrm{R}$ and symmetric polynomials in $\mathscr{R}$:
\begin{align}   \label{corrhwR} 
 \sum_\La c_\La G_{-\La^a}L_{-\La^s}|h\R^{\mathrm{R}}& \longleftrightarrow \sum_\La c_\La \mathcal{G}_{-\La^a}\mathcal{L}_{-\La^s}(1)
\end{align}
where 
\beq h=\frac{1}{2}{\eta}(\eta-2\gamma)+\frac{1}{16}. \eeq

\subsection{Singular vectors} 

{As in the NS sector, the singular vectors  are captured in a remarkably simple way in the sJack-basis.} 
To apply the correspondence \eqref{corrhwR}  to the $(r,s)$-{type} Kac module, we must further set
\beq t=\alpha,\qquad\gamma=\frac{1}{2\sqrt{\alpha}}(\alpha-1) \qquad \text{and}\qquad \eta\,
{\equiv \eta_{r,s}}=\frac{1}{2\sqrt{\alpha}}\left((r+1)\alpha-(1+s)\right) \,.\eeq
According to \eqref{condsingvecR}, the bosonic singular vector $|\chi_{r,s}\R^\mathrm{R}$ is {represented by} the superpolynomial ${F}_{r,s}$ if, and only if, 
 \beq  {\mathcal{G}_1(F_{r,s})=0\qquad \text{and}\qquad  \mathcal{L}_1(F_{r,s})=0 \,  }
\eeq
where \beq \label{expandR} {F_{r,s}=\sum_{\La\in\mathscr{P}_{r,s}}c_\La P_\La^{(\alpha)}\, , }\eeq 
and $\mathscr{P}_{r,s}$ denotes the set of superpartitions $\La$ such that $\deg (\La)=(\frac{rs}2|k)$ with $k=0 \mod 2$.
The fermionic partner of $|\chi_{r,s}\R^\mathrm{R}$ is simply represented by $\mathcal{G}_0(F_{r,s})$.

Let us display a few explicit formulas.  We choose the normalization in \eqref{expandR} to be such that $c_{(r^{s/2})}=1$
and  $c_{(r^{(s-1)/2},r/2)}=1$ when $r$ is odd or even, respectively.    For $r=1$ and $s$ even,  we find
\beq 
\boxed{ |\chi_{1,s}\R^\mathrm{R} \longleftrightarrow P^{\alpha}_{( 1^{{s}/{2}})}- P^{\alpha}_{(1,0; 1^{s/{2}-1})} }
\eeq
while its  dual expression reads
\beq 
\boxed{ |\chi_{r,1}\R^\mathrm{R} \longleftrightarrow P^{\alpha}_{( \frac{r}{2})}+\frac{\alpha r}{2} P^{\alpha}_{(\frac{r}{2},0;)} .}
\eeq
The $r=2$ case is also very simple:
\beq \boxed{|\chi_{2,s}\R\longleftrightarrow (\alpha+1)\sum_{\ell=0}^{(s-1)/2} \prod_{i=\ell+1}^{2\ell+1}\frac{i}{(\a+i)} \, \left(P^{(\a)}_{(2^{(s-1)/2-\ell}1^{2\ell+1})}+\frac{\alpha}{2\ell+1} \,P^{(\a)}_{(1,0;2^{(s-1)/2-\ell}1^{2\ell})} \right).
}\eeq

 From these examples, we see that the self-complementary rule  
 needs to be slightly
modified in the R sector. Indeed, we now have that a superpartition $\Lambda$
is $(r,s)$-self-complementary in the R sector if the complement of
$\Lambda^*$ in the rectangle with $r$ columns and $s$ rows corresponds to
$\Lambda^*$ rotated by $\pi$. 
This modified rule has been verified to high degree.
For instance, here are the splittings of the rectangles with $(r,s)=(1,6)$ and $(2,5)$
(with the boxes of the rotated copy of $\La^*$ marked with thick frames):
\beq
\mbox{\small
(1,6):\;
\tableau[scY]{\\ \\ \\\tf\\ \tf\\ \tf}\;\qquad\qquad
(2,5):\;
\tableau[scY]{&\\&\\&\tf\\\tf&\tf\\\tf&\tf}\qquad
\tableau[scY]{&\\&\tf\\&\tf\\ &\tf\\\tf&\tf}\qquad
\tableau[scY]{&\tf\\&\tf\\&\tf\\&\tf\\ &\tf} \;.
}
\eeq
In the first case, the only possible value of $\La^*$ is $(1^3)$. But this can be associated to two superpartitions: $(1^3)$ and $(1,0;1,1)$. No higher value of the fermionic degree can be realized. Similarly, each $(2,5)$-diagram can be realized by two superpartitions, respectively: $(2^2,1)$ and $(1,0;2^2)$, $(2,1^3)$ and $ (1,0;2,1)$ and finally, $(1^5)$ and $(1,0;1^3)$. For $r>2$, there are typically more than two superpartitions associated to each splitting of the rectangle. 
Consider for instance the case $(r,s)=(3,4)$; the contributing diagrams and the corresponding superpartitions are
\beq
\tableau[scY]{&&\\&&\\\tf&\tf&\tf\\\tf&\tf&\tf\\}
\longleftrightarrow\;
\footnotesize{ \begin{matrix}(3,3)\\(3,0;3)\end{matrix}}\qquad
\normalsize{\tableau[scY]{&&\\&&\tf\\&\tf&\tf\\\tf&\tf&\tf\\}
\longleftrightarrow}\;
\footnotesize{ \begin{matrix}(3,2,1)\\(1,0;3,2)\\(2,0;3,1)\\(2,1;3)\\(3,0;2,1)\\(3,1;2)\\(3,2;1)\\(3,2,1,0;)\end{matrix} } \qquad \normalsize{
\tableau[scY]{&&\tf\\&&\tf\\&\tf &\tf\\&\tf&\tf\\}
\longleftrightarrow }\;
\footnotesize{ \begin{matrix}(2,2,1,1)\\(2,1;2,1)\end{matrix}}\qquad
\eeq

\subsection{Degenerate Whittaker {vector}}
The NS degenerate Whittaker vector, found to be simple in the sJack basis, also  has a simple representation in the R sector. Again, we}
work within  the highest-weight module of positive chirality characterized by the  highest weight $|h\R^\mathrm{R}$.  We define the bosonic component of the degenerate Whittaker vector at level $k$ as the state $|h\R^\mathrm{R}_k$ {defined by the conditions} \cite{Ito} : 
\beq \label{whittR}    L_{1} |h\R_k^\mathrm{R}=   |h\R_{k-1}^\mathrm{R},\qquad G_{1}  |h\R_k^\mathrm{R}=0, \qquad \forall\, k\geq 1\, ,
\eeq
with the initial condition $|h\R_0^\mathrm{R}=|h\R^\mathrm{R}$. We have found that these vectors can be represented in the sJack basis as follows:
\beq \boxed{ |h\R_k^\mathrm{R}\longleftrightarrow 
\tilde W_k=2^k\!\!\sum_{\substack{\La,\,|\La|=k\\ m=0\, \text{mod}\, 2}}
\a^{|\Lambda^\cd|} \,\tilde w_\La P_\La} \eeq
and 
\beq\label{wRs}\boxed{
\tilde w_\La=  
 \left[\prod_{s\in \B\La}\frac1{h^\uparrow_\La(s)}\right]\, 
\, \left[\frac{ \prod_{(i,j)\in \B\La}[2\bar\eta+i+\la_j'+\e_j'-\a(j+\la_i+\e_i)]}{  
\prod_{(i,j)\in\La} [2\bar\eta+2i-\a(2j+1)]\,[2\bar\eta+2i+1-\a2j]} \right] }
\eeq with, as before,  $\bar\eta=\sqrt{\a}\, \eta$,  
{and where  $\e_i=0$ if the $i$-th row is fermionic and 1 otherwise and $\e'_j=0$ if the $j$-th column is fermionic and 1 otherwise. Note that 
$\B\La$ was defined after \eqref{defA}}.

Let us illustrate the evaluation of the second square bracket in \eqref{wRs}, denoted $\tilde A$ below,  by displaying some examples using the compact notation
$(2\bar\eta+a-\a b)\equiv(a,b)$. 
We have 
\beq
{\tableau[scY]{&&\bl\tcercle{}\\ \bl\tcercle{}\\ }}
\;\rw\;\tilde A=\frac1{(3,2)\,(2,3)\,(2,5)};,\qquad
{\tableau[scY]{&& \bl\tcercle{}\\ & \bl\tcercle{}\\ }}\;\rw\;\tilde A=\frac1{(2,3)\,(2,5)\,(3,2)\,(3,4)},
\eeq and 
\beq
{\tableau[scY]{&\\& \\ }}
\;\rw\; \tilde A=\frac{(4,4)\,(4,5)\,(5,4)\,(5,5)}{(2,3)\,(2,5)\,(4,3)\,(4,5)\,(3,2)\,(3,4)\,(5,2)\,(5,4)}.
\eeq
{
Finally, specializing the general  expression of $\ti W_k$ to $k=2$}, we get, 
\begin{align}
\tilde W_2 &= {\frac {2 (3,5) }{
 ( 2,3)  (2,5 )  ( 3,2  ) }}P^{(\a)}_{{(;2)}}+{\frac {
4 \a\,( 5,3 )  }{
 ( \a +1 )  (3,2 ) 
 ( 2,3 )  (5,2) }}P^{(\a)}_{{(;1,1)}}
 \nonumber\\ &+{\frac {4\a }{ ( 2,3)  (2,5 ) 
 (3,2  ) }} P^{(\a)}_{{(2,0;)}}
 +{\frac {4{\a }^{2}}{ ( \a +1 )  ( 3,2  )  ( 2,3)  (5,2) }}P^{(\a)}_{{(1,0;1)}}
.
 \end{align}

\section{Tables of Jack superpolynomials}\label{TablesJacks}
\newcolumntype{C}{>{$\displaystyle}c<{$}}
\newcolumntype{R}{>{$\displaystyle}r<{$}}
\newcolumntype{L}{>{$\displaystyle}l<{$}}

\subsection{sJacks in terms of monomials}

We are interested in the transition matrices $c_{\La\Om}$ relating the sJack basis to the supermonomial basis, i.e.,  $P_\La=\sum_\Om c_{\La\Om} m_\Om$.  
{To make the presentation more compact}, we eliminate the parentheses and the commas of the   superpartitions, so that, for instance, $(1,0;3,2)$ {becomes} $10;32$. 
{In the following tables, the superpartitions labeling the rows are the $P^{(\a)}$-labels and the row-entries are the coefficients of the monomials labeled by the partition specifying the column. For instance,} the simplest nontrivial transition matrix and its corresponding explicit formulas read
{\small{ \begin{center} \begin{tabular}{r|cc} 
{$P\backslash m$}\! \!\!\!\!& $1;$ & $0;1$ \\ \hline  
1;& $1  $& $\displaystyle {(\alpha+1)^{-1} }$\\ 
 0;1 & $0   $ & 1 \\ 
\end{tabular} $\displaystyle \qquad\Longleftrightarrow\qquad  P^{(\a)}_{1;}=m_{1;}+\frac{1}{ \alpha+1 }m_{0;1}, \qquad  \text{and}\qquad P^{(\a)}_{0;1}=m_{0;1}\, .  $ \end{center} }}\normalsize

Below we display all transition matrices for  superpartitions of degree $(n|m)$ with $n,m\leq 3$.  For $n=2$, we have 
 {\small \begin{center} 
\begin{tabular}{r|cc}
    $P\backslash m\!\!\!$ &2&11\\ \hline 
2  & 1  &$\displaystyle {2}({\alpha+1})^{-1}$\\  
11 & 0 & 1\\  
\end{tabular}  
\hspace{1cm}
\begin{tabular}{R|CC}
 P\backslash m  \!\!\!\!  &20; & 10;1  \\ \hline 
 20;  & 1 & ({\alpha+1})^{-1}\\  
10;1& 0 & 1\\  
\end{tabular}
\end{center} }\normalsize
\noindent and 
 {\small \begin{center} 
\begin{tabular}{R|CCCC}  P\backslash m \!\!\!\! &2; &0;2&1;1& 0;11\\ 
\hline
 2; &1& \left( 1+2\,\alpha \right) ^{-1}&2 \left( 1+2\,\alpha \right) ^{-1}&{ {2} { \left( \alpha+1
 \right)^{-1}  \left( 1+2\,\alpha \right)^{-1} }}\\ 
 0;2&0
&1& \left( \alpha+1 \right) ^{-1}&2 \left( \alpha+1 \right) ^{-1}
\\ 
 1;1 &0&0&1&2\left( 2+\alpha \right) ^{-1}
\\  
0;11&0&0&0&1\end{tabular}
\end{center} }\normalsize
\noindent For $n=3$, apart from $P_{210;}=m_{210;}$, we have \\ 
{\footnotesize  \begin{tabular}{R|CCC}  P\backslash m  \!\!\!\!  & 3 & 21 & 111  \\ \hline 3 & 1& \frac{3}{1+2\,\alpha } & \frac{6}{ \left( \alpha+1 \right)  \left( 1+2\,\alpha \right) } \\ 21 &  0&1& \frac{6}{ 2+\alpha}  \\ 111 &  0 &0&1  \end{tabular} 
\hspace{.5cm} 
\begin{tabular}{R|CCCCC}  P\backslash m  \!\!\!\! &30;&21;&20;1&10;2&10;11\\ \hline
30; &1& \frac{1}{ 1+2
\alpha }&\frac{2}{ 1+2\alpha }& \frac{1}{ 1+2\alpha  }&{\frac {2}{ \left( \alpha+1 \right)  \left( 
1+2\alpha \right) }}\\
21;&0&1& \frac{1}{
\alpha+1 }&-{\frac {\alpha}{ 2\left( \alpha+1 \right) 
^{2}}}& \frac{1}{\left( \alpha+1 \right) ^{2}}\\
20;1&0
&0&1& \frac{1}{ \alpha+1 }&\frac{2}{\alpha+1 }
\\ 
10;2&0&0&0&1& \frac{2}{2+\alpha}\\ 
10;11&0&0&0&0&1
\end {tabular} 
}\normalsize
and 
{\footnotesize \begin{center} 
\begin{tabular}{R|CCCCCCC} P\backslash m  \!\!\!\! &3;&0;3&2;1&1;2&0;21&1;11&0;111\\\hline
3;
&1& \!\! \frac{1}{ 1+3\,\alpha}\!\! & \!\! \frac{3}{ 1+3\,\alpha} \!\! &\!\! {\frac {3(\alpha+1)}{ \left( 1+2\alpha \right)  \left( 1+3
\alpha \right) }}\!\! &\!\! {\frac {3}{ \left( 1+2\alpha \right)  \left( 1+
3\alpha \right) }}\!\! & \!\! {\frac {6}{ \left( 1+2\alpha \right) 
 \left( 1+3\alpha \right) }} \!\! &\!\! {\frac {6}{ \left( 1+2\alpha
 \right)  \left( 1+3\alpha \right)  \left( \alpha+1 \right) }} \!\!
\\ 
0;3&0&1& \!\! \frac{1}{ 1+2\,\alpha }\!\! &\!\! \frac{2}{ 1+2\,\alpha } \!\!& \!\! \frac{3}{ 1+2\,\alpha}\!\! &{\frac {4}{ \left( \alpha+1 \right)  \left( 1+2\alpha \right) 
}}&{\frac {6}{ \left( \alpha+1 \right)  \left( 1+2\alpha \right) 
}}\\
2;1&0&0&1&\frac{1}{ \alpha+1}&
{\frac {2+\alpha}{ 2\left( \alpha+1 \right) ^{2}}}&{\frac {2
\alpha+3}{ \left( \alpha+1 \right) ^{2}}}&\frac{3}{\left( \alpha+1 \right) 
^{2}}\\
1;2&0&0&0&1& \frac{1}{\alpha+1}&\frac{2}{ \alpha+1}&{\frac {6}{ \left( 2+\alpha
 \right)  \left( \alpha+1 \right) }}\\ 
0;21&0&0&0&0&1&\frac{2}{ 2+\alpha}&\frac{6}{ 2+\alpha}\\ 
1;11 &0&0&0&0&0&1&\frac{3}{ 
3+\alpha }\\0;111&0&0&0&0&0&0&1
\end{tabular}
\end{center}
}\normalsize

\subsection{sJacks in terms of power-sums}

We now write down some transition matrices $e_{\La,\Om}$ relating the sJack basis to that constructed with the power-sums, i.e.,  $P_\La=\sum_\Om e_{\La\Om} p_\Om$.  
The simplest nontrivial transition matrix is found at degree $(1|0)$ {corresponding to eqs \eqref{Pex2} and \eqref{Pex1})}:
{\small { \begin{center} \begin{tabular}{r|cc} 
{$P\backslash p$}\!\!& $1;$ & $0;1$ \\ \hline  
$\displaystyle \!\!\phantom{\frac{|}{|}} 1$;& $\displaystyle \frac{\alpha}{\alpha+1} $& $\displaystyle \frac{1}{\alpha+1}$\\ 
 0;1 & $-1  $ & 1 \\ 
\end{tabular}     \end{center} }}\normalsize
\noindent For $n=2$, we have 
 {\small \begin{center} 
\begin{tabular}{r|cc}
    $P\backslash p\!\!\!$ &2&11\\ \hline 
$\displaystyle \!\!\phantom{\frac{|}{|}} \!\! 2 $ & $\displaystyle \frac{\alpha}{\alpha+1}$ &$\displaystyle \frac{1}{\alpha+1}$\\  
$\displaystyle \!\!\phantom{\frac{|}{|}} \!\! 11 $& $\displaystyle -\frac{1}{2}$ & $\displaystyle \frac{1}{2}$\\  
\end{tabular}  
\hspace{1cm}
\begin{tabular}{R|CC}
  P\backslash p \!\!\!\!  &20; & 10;1  \\ \hline 
\!\!\phantom{\frac{|}{|}}\!\! 2,0;  & \frac{\alpha}{\alpha+1} & \frac{1}{\alpha+1}\\  
\!\!\phantom{\frac{|}{|}} \!\!10;1& -1 & 1\\  
\end{tabular}
\end{center} }\normalsize
\noindent and 
 {\small \begin{center} 
\begin{tabular}{R|CCCC} P\backslash p \!\!\!\! &2; &0;2&1;1& 0;11\\ 
\hline
 2; &2\alpha^2(\alpha+1)^{-1}(2\alpha+1)^{-1}& \alpha(\alpha+1)^{-1}(2\alpha+1)^{-1}&2\alpha(\alpha+1)^{-1}(2\alpha+1)^{-1}& (\alpha+1)^{-1}(2\alpha+1)^{-1}\\ 
 0;2&-\alpha(\alpha+1)^{-1}& \alpha(\alpha+1)^{-1}& -(\alpha+1)^{-1}& \left( \alpha+1 \right) ^{-1}
\\ 
 1;1 &-\alpha(\alpha+2)^{-1}& -(\alpha+2)^{-1}& \alpha (\alpha+2)^{-1}& \left( \alpha+2 \right) ^{-1}
\\  
0;11&1&-2^{-1}&-1&2^{-1}\end{tabular}
\end{center} }\normalsize 
\noindent For $n=3$, we have the trivial case $P_{210;}=p_{210;}$ and \\
{\footnotesize \noindent \begin{tabular}{R|CCC}  P\backslash p  \!\!\!\!  & 3 & 21 & 111  \\ \hline
 3 & \!\!\!  \frac{2\alpha^2}{[1,1][2,1]} \!\!\! & \!\!\! \frac{3\alpha}{[1,1][2,1]} \!\!\!  & \!\!\!  {\frac{1}{ [1,1][2,1] }} \!\!\! \\ 
21 & -
\frac{\alpha}{ [1,2] }&\frac{[1,-1]}{[1,2]}& \frac{1}{[1,2]} \\ 
111 &  \frac{1}{3} &-\frac{1}{2}&\frac{1}{6}  \end{tabular} 
\hspace{.5cm} 
\begin{tabular}{R|CCCCC}  P\backslash p  \!\!\!\! &30;&21;&20;1&10;2&10;11\\ \hline
30; &   \!\!\!  \frac{2\alpha^2}{[1,1][2,1]} \!\!\!  &  0 &  \!\!\!  \frac{2\alpha}{[1,1][2,1]} \!\!\!  &   \!\!\!  \frac{\alpha}{[1,1][2,1]} \!\!\!  &   \!\!\!  \frac{1}{[1,1][2,1]} \!\!\!  \\
21;& \frac{-\alpha}{2[1,1]^2} & \frac{\alpha[2,1]}{2[1,1]^2}& \frac{\alpha}{[1,1]^2}&\frac {-\alpha}{ 2[1,1]}& \frac{1}{2[1,1]^2}\\
20;1& \frac{-\alpha}{[1,1]} &  \frac{-\alpha}{[1,1]}  & \frac{[1,-1]}{[1,\phantom{-}1]}  & 0&\frac{1}{[1,1]}
\\ 
10;2&\frac{-\alpha}{[1,2]} & 1  & \frac{-2}{[1,2]}  & \frac{[1,1]}{[1,2]}&\frac{1}{[1,2]}
\\ 
10;11&1&0&-1&-\frac{1}{2}&\frac{1}{2}
\end {tabular} 
}\normalsize\\
where we have used the following shorthand notation: 
\beq [a,b]\,\equiv\, a\,\alpha +b \,.\eeq
Finally,
{\footnotesize \noindent \begin{center} 
\begin{tabular}{R|CCCCCCC} P\backslash p  \!\!\!\! &3;&0;3&2;1&1;2&0;21&1;11&0;111\\\hline
3;&      \!\!\!  \frac{6\alpha^3}{[1,1][2,1][3,1]} \!\!\!  & \!\!\!  \frac{2\alpha^2}{[1,1][2,1][3,1]} \!\!\!  & \!\!\!  \frac{6\alpha^2}{[1,1][2,1][3,1]} \!\!\!  &\!\!\!  \frac{3\alpha^2}{[1,1][2,1][3,1]} \!\!\! &\!\!\!  \frac{3\alpha}{[1,1][2,1][3,1]} \!\!\! &\!\!\!  \frac{3\alpha}{[1,1][2,1][3,1]} \!\!\! &
\!\!\!  \frac{1}{[1,1][2,1][3,1]} \!\!\! 
\\ 
0;3& \frac{-2\alpha^2}{[1,1][2,1]} &\frac{2\alpha^2}{[1,1][2,1]} & \frac{-2\alpha}{[1,1][2,1]} & \frac{-\alpha}{[1,1][2,1]} & \frac{3\alpha}{[1,1][2,1]} & \frac{-1}{[1,1][2,1]} & \frac{1}{[1,1][2,1]}  \\
2;1& \frac{-\alpha^2}{[1,1]^2} & \frac{-\alpha}{2[1,1]^2} & \frac{\alpha[2,-1]}{2[1,1]^2} & \frac{-\alpha }{2[1,1]^2} & \frac{[1,-1]}{2[1,1]^2} &  \frac{\alpha}{[1,1]^2} & \frac{1}{2[1,1]^2}  
\\
1;2& \frac{-\alpha[1,-1]}{[1,1][1,2]} &  \frac{-\alpha }{[1,1][1,2]} &  \frac{-3\alpha}{[1,1][1,2]} &  \frac{\alpha^2+\alpha+1}{[1,1][1,2]}  &  \frac{[1,-1]}{[1,1][1,2]} &
 \frac{[1,-1]}{[1,1][1,2]} &  \frac{1}{[1,1][1,2]}  
\\ 
0;21& \frac{2\alpha}{[1,2]} & \frac{-\alpha}{[1,2]} & \frac{[-1,2]}{[\phantom{-}1,2]} & \frac{-\alpha}{[1,2]} &
\frac{[1,-1]}{[1,\phantom{-}2]} &  \frac{-2}{[1,2]} & \frac{1}{[1,2]} 
\\ 
1;11 & \frac{\alpha}{[1,3]}&  \frac{1}{[1,3]}& \frac{-\alpha}{[1,3]}& \frac{-\alpha}{2[1,3]}& \frac{-3}{2[1,3]}& \frac{\alpha}{2[1,3]}& \frac{1}{2[1,3]}
\\
0;111& -1 &\frac{1}{3} & 1 &\frac{1}{2} & -\frac{1}{2} & -\frac{1}{2} &\frac{1}{6} 
\end{tabular}
\end{center}
}

\end{appendix}


\begin{thebibliography}{99}
\addcontentsline{toc}{section}{References}

\bibitem{Alba}
V. A. Alba, V.A. Fateev, A. V. Litvinov and G. M. Tarnopolskiy,
{\it On Combinatorial Expansion of the Conformal Blocks Arising from AGT Conjecture},
Lett. Math. Phys. {\bf 98} (2011) 33-64. 

\bibitem{AGT}
L.\ F. Alday, D. Gaiotto and Y. Tachikawa,
{\it Liouville correlation functions from four-dimensional gauge theories}
Lett.\ Math.\ Phys.\ {\bf 91} (2010) 167--197.


 \bibitem{AMOSa} H. Awata, Y. Matsuo, S. Odake and J. Shiraishi, {\it Collective field theory, Calogero- Sutherland model and generalized matrix models}, Phys.\ Lett.\ {\bf B347} (1995) 49--55;  {\it Excited states of the Calogero- Sutherland model and singular vectors of the $W_N$ algebra}, Nucl. Phys. {\bf B449} (1995) 347--374.
	
	
	
\bibitem{AY} H.\ Awata and Y.\ Yamada, {\it Five-dimensional AGT conjecture and the deformed Virasoro algebra}, JHEP01 (2010) 125, 1--11.	
	

\bibitem{BB}
A. Belavin, V. Belavin,
{\it AGT conjecture and Integrable structure of Conformal field theory for $c=1$},
Nucl. Phys. {\bf B850} (2011) 199-213.


\bibitem{BF}
V. Belavin and B. Feigin, 
{\it Super Liouville conformal blocks from $N=2$ $ SU(2)$ quiver gauge theories}, JHEP 07 (2011) 079. 


\bibitem{BBB}
A. Belavin, V. Belavin and M. Bershtein, {\it Instantons and 2d superconformal field theory}
JHEP 09 (2011) 117.

{
\bibitem{coset}
A.A. Belavin, M.A. Bershtein, B.L. Feigin, A.V. Litvinov, G.M. Tarnopolsky,
{\it Instanton moduli spaces and bases in coset conformal field theory},
 arXiv:1111.2803}


\bibitem{BsA}
L.\ Benoit and Y.\ Saint-Aubin, {\it Singular vectors of the Neveu-Schwarz algebra},
Int.\ J.\ of Mod.\ Phys.\ A {\bf 7} (1992) 3023--3033.


\bibitem{BKT}
M. Bershadsky, V. Knizhnik, M. Teitelman, {\it Superconformal symmetry in two dimensions}
Phys. Lett., {\bf 151B} (1985) 31-36.



\bibitem{BMTa}
G. Bonelli, K. Maruyoshi and A. Tanzini, {\it Instantons on ALE spaces and Super Liouville Conformal Field Theories}, arXiv:1106.2505; 
G. Bonelli, K. Maruyoshi, and A. Tanzini, {\it Gauge Theories on ALE Space and Super Liouville Correlation Functions},
arXiv:1107.4609.

\bibitem{BTW}	
L.\ Brink, A.\ Turbiner, and N.\ Wyllard, {\it Hidden algebras of the (super) Calogero and Sutherland models}, J.\ Math.\ Phys.\ 39 (1998) 1285--1315.


\bibitem{CJ}
W. Cai and N. Jing, {\it Applications of Laplace-Beltrami operator for Jack polynomials}, arXiv:1101.5544v2.


  \bibitem{DLM1}
P.~Desrosiers, L.~Lapointe and P.~Mathieu, \emph{Supersymmetric
Calogero-Moser-Sutherland models and Jack superpolynomials}, Nucl.\
Phys.\ {\bf B606} (2001)  547--582.




\bibitem{DLM3}
P.~Desrosiers, L.~Lapointe and P.~Mathieu, \emph{Jack polynomials in
superspace}, Commun.\ Math.\ Phys.\ {\bf 242} (2003)  331--360.


\bibitem{DLM6}
P.~Desrosiers, L.~Lapointe and P.~Mathieu, \emph{Classical symmetric functions in  superspace}, J. Alg. Comb., {\bf 24} (2006) 209-238.



\bibitem{DLM7}
P.~Desrosiers, L.~Lapointe and P.~Mathieu, \emph{Orthogonality of Jack polynomials in superspace}, Adv. Math.{\bf 212} (2007) 361--388.

\bibitem{DLMeva}
P.~Desrosiers, L.~Lapointe and P.~Mathieu, \emph{Evaluation and normalization of Jack polynomials in superspace}, arXiv:1104.3260, to appear in Int. Math. Res. Not.




\bibitem{DLM0}
P.~Desrosiers, L.~Lapointe and P.~Mathieu,
{\it Jack superpolynomials with negative fractional parameter: clustering properties and super-Virasoro ideals}, arXiv:1109.2832.


 \bibitem{CFT}
P.~Di Francesco, P. Mathieu, and D. S\'{e}n\'{e}chal,
{\em {Conformal Field Theory}},
Springer-Verlag, New York, 1997. 
 

\bibitem{Dorr} M.\  Dorrzapf, \textit{ Highest weight representations of the $N= 1$ Ramond algebra}, Nucl.\ Phys.\ B {\bf 595} (2001) 605-653.


\bibitem{DF}
V.S.  Dotsenko and V.A. Fateev, {\it Conformal algebra and multipoint correlation functions
in 2d statistical models}, Nucl. Phys. {\bf B 240} (1984) 312-348.


\bibitem{FQS}
D. Friedan, Z. Qiu, S.H. Shenker, {\it nce in two dimensions and the tricritical Ising model},
Phys. Rev. Lett., {\bf 151B} (1985) 37-43.

\bibitem{Gai}
D. Gaiotto,
{\it Asymptotically free N=2 theories and irregular conformal blocks}
arXiv:0908.0307 [hep-th]


\bibitem{KW} V.\ G.\ Kac and M.\ Wakimoto, {\it Modular Invariant Representations of Infinite-Dimensional Lie Algebras and Superalgebras}, Proc.\ Nat.\ Acad.\ Sci.\ USA {\bf 85} (1988) 4956--4960.

\bibitem{Kirillov}  A.\ A.\ Kirillov (Edt), {\it Representation Theory and Noncommutative Harmonic Analysis I}, Springer-Verlag, Berlin Heidelberg, 1994.

\bibitem{IK1}
K.\ Iohara and Y.\ Koga, {\it Representation theory of Neveu-Schwarz and
Ramond algebras I: Verma modules}, Adv.\ Math.  {\bf 178} (2003) 1--65; {\it Representation theory of Neveu-Schwarz and Ramond Algebras II: Fock modules}, Ann.\ Inst.\  Fourier {\bf 53} (2003) 1755--1818.

\bibitem{IK2}
K.\ Iohara and Y.\ Koga, {\it Singular Vectors of the $N=1$ Superconformal Algebra}, Ann.\ Henri Poincar\'e {\bf 3} (2002) 19--27.

\bibitem{Ito}
Y. Ito,
{\it Ramond sector of super Liouville theory from instantons on an ALE space}, arXiv:1110.2176, 20 pages.




\bibitem{LLN}
L. Lapointe, Y. Le Borgne and P. Nadeau
\emph{A normalization formula for the Jack polynomials in superspace and an identity on partitions},
Electronic J. Comb.  {\bf 16} (2009) Article \#R70.



\bibitem{Mac}
               {I.~G.~ Macdonald},
                \emph{Symmetric functions and {H}all polynomials},
2nd ed., The Clarendon Press/Oxford University Press
                (1995).
                 
                 


\bibitem{MY}
 K. Mimachi and Y. Yamada, {\it Singular vectors of the Virasoro algebra in terms of Jack symmetric polynomials}, Comm. Math. Phys. {\bf174} (1995) 447-455.
 

\bibitem{MMS}A.\ Mironov, A.\ Morozov, and Sh.\ Shakirov, \textit{A direct proof of AGT conjecture at $\beta$ = 1}, JHEP02 (2011), 1--40; A.\ Mironov, A.\ Morozov, Sh.\ Shakirov, and A. Smirnov, \textit{
Proving AGT conjecture as HS duality: Extension to five dimensions}, Nucl.\ Phys.  B {\bf 855}, 128--151.



{\bibitem{Nek}N. A. Nekrasov, {\it Seiberg-Witten Prepotential from Instanton Counting}, Adv. Theor. Math. Phys. {\bf 7} (2003) 831-864.}

\bibitem{Segal} G.\ Segal, \textit{Unitary representations of some infinite dimensional groups}, Commun.\ Math.\ Phys.\ 80 (1981) 301--342.

\bibitem{SSAFR}
 R. Sakamoto, J. Shiraishi, D. Arnaudon, L. Frappat, and E. Ragoucy, {\it Correspondence between conformal field theory and Calogero-Sutherland model}, Nucl. Phys. {\bf B 704} (2005) 490-509. 
 

    
\bibitem{SS}
B.~S. Shastry and B.~Sutherland, \emph{Superlax pairs and infinite
symmetries in the $1/r^2$ system}, Phys.\ Rev.\ Lett.\ {\bf 70}
(1993) 4029--4033;

\bibitem{Stan}
               R.~P.~Stanley,
               \emph{Some combinatorial properties of Jack symmetric
functions}, Adv.\ Math.\ {\bf77} (1988) 76--115.


 
 \bibitem{TK}
 A. Tsuchiya and Y. Kanie, {\it Fock space representation of the Virasoro algebra --- Intertwining operators}, Publ. RIMS Kyoto Univ. {\bf 22} (1986) 259--327.


\bibitem{WY} M.\ Wakimoto and H.\ Yamada, \textit{Irreducible decompositions of Fock representations of
the Virasoro algebra}, Lett.\ Math.\ Phys.\ {\bf 7} (1983) 513--516; \textit{The Fock representations of the Virasoro algebra and the Hirota equations of the modified KP hierarchies}, Hiroshima Math.\ J.\ {\bf 16} (1986) 427--441.
 

\bibitem{Wal} N.\ R.\ Wallach, \textit{Classical invariant theory and the Virasoro algebra}, in "Vertex operators
in Mathematics and Physics", J. Lepowsky et al. ed., Springer, 1985, 475--482.


\bibitem{Watts} G.\ M.\ T.\  Watts, \textit{Null vectors of the superconformal algebra: The Ramond sector}, 
Nucl.\ Phys.\ B {\bf 407}  213--236.

 \bibitem{Yan}
S. Yanagida,
{\it Whittaker vectors of the Virasoro algebra in terms of Jack symmetric polynomial}, J.\ Alg.\ {\bf 333} (2011) 273--294.
 

\end{thebibliography}
\end{document}